\documentclass[twocolumn]{aastex701}
\usepackage{times}
\usepackage{amsmath}
\usepackage{textcomp}
\usepackage{amssymb}
\usepackage{lineno}
\usepackage{gensymb}
\usepackage{hyperref}
\usepackage{url}
\usepackage{afterpage}
\usepackage{float}
\usepackage{mathrsfs}
\usepackage{rotating}
\usepackage{csquotes}
\usepackage{graphicx}
\usepackage[caption=false]{subfig}
\usepackage{booktabs}

\newcommand{\fermi}{{\it Fermi}}

\newcommand{\rad}{$R_{95}$}

\shorttitle{2FLE}
\shortauthors{{\it Fermi}-LAT collaboration}

\begin{document}

\title{{\large The Second Catalog of Low-Energy \textit{Fermi}-LAT Sources}}

\author[0009-0004-3923-9884]{A.~Adelfio}
\affil{Istituto Nazionale di Fisica Nucleare, Sezione di Perugia, I-06123 Perugia, Italy}
\email{}
\author[0000-0002-6584-1703]{M.~Ajello}
\email{majello@clemson.edu}
\affil{Department of Physics and Astronomy, Clemson University, Kinard Lab of Physics, Clemson, SC 29634-0978, USA}
\author[0009-0001-2927-8968]{E.~Aviano}
\affil{Dipartimento di Fisica, Universit\`a di Trieste, I-34127 Trieste, Italy}
\affil{Istituto Nazionale di Fisica Nucleare, Sezione di Trieste, I-34127 Trieste, Italy}
\email{}
\author[0000-0002-9785-7726]{L.~Baldini}
\affil{Universit\`a di Pisa, Dipartimento di Fisica E. Fermi, I-56127 Pisa, Italy}
\affil{Istituto Nazionale di Fisica Nucleare, Sezione di Pisa, I-56127 Pisa, Italy}
\email{}
\author[0000-0002-8784-2977]{J.~Ballet}
\affil{Universit\'e Paris-Saclay, Universit\'e Paris Cit\'e, CEA, CNRS, AIM, F-91191 Gif-sur-Yvette Cedex, France}
\email{}
\author[0000-0001-7233-9546]{C.~Bartolini}
\affil{Istituto Nazionale di Fisica Nucleare, Sezione di Bari, I-70126 Bari, Italy}
\affil{Universit\`a degli studi di Trento, via Calepina 14, 38122 Trento, Italy}
\email{}
\author[0000-0002-6954-8862]{D.~Bastieri}
\affil{Istituto Nazionale di Fisica Nucleare, Sezione di Padova, I-35131 Padova, Italy}
\affil{Dipartimento di Fisica e Astronomia ``G. Galilei'', Universit\`a di Padova, Via F. Marzolo, 8, I-35131 Padova, Italy}
\affil{Center for Space Studies and Activities ``G. Colombo", University of Padova, Via Venezia 15, I-35131 Padova, Italy}
\email{}
\author[0000-0002-6729-9022]{J.~Becerra~Gonzalez}
\affil{Instituto de Astrof\'isica de Canarias and Universidad de La Laguna, Dpto. Astrof\'isica, 38200 La Laguna, Tenerife, Spain}
\email{}
\author[0000-0002-2469-7063]{R.~Bellazzini}
\affil{Istituto Nazionale di Fisica Nucleare, Sezione di Pisa, I-56127 Pisa, Italy}
\email{}
\author[0000-0001-9935-8106]{E.~Bissaldi}
\affil{Dipartimento di Fisica ``M. Merlin" dell'Universit\`a e del Politecnico di Bari, via Amendola 173, I-70126 Bari, Italy}
\affil{Istituto Nazionale di Fisica Nucleare, Sezione di Bari, I-70126 Bari, Italy}
\email{}
% ...

\begin{abstract}
We used 14 years of \textit{Fermi}-Large Area Telescope (LAT) data to derive a catalog of sources detected between 20 MeV and 200 MeV. For each source we provide localization, spectral, and association information. The Second Catalog of Low-Energy \textit{Fermi}-LAT sources (2FLE) contains 450 sources detected at $TS>25$, with median energy flux of $F_{20-200 \, \text{MeV}} = 1.9\times10^{-11}$ erg cm$^{-2}$ s$^{-1}$. 
Using a Bayesian and likelihood-ratio method, we associate 377 2FLE sources with a known counterpart. Of the 325 objects associated with extragalactic sources, 
the dominant source class represented in the 2FLE are flat-spectrum radio quasars (FSRQ). Additionally, 52 sources are associated with Galactic sources. Three associated sources have not been previously detected in any $\gamma$-ray catalog and represent new $\gamma$-ray sources of known origin. 
We find that 387 2FLE sources have a positional cross-match with the 4FGL-DR4 sources. For the sources with a 4FGL counterpart, we compare the spectrum of the 2FLE source to the 4FGL spectral energy distributions.
Of the 73 2FLE sources without multi-wavelength associations,
13 have a 4FGL counterpart that lacks classification, while 60 are newly discovered $\gamma$-ray sources. Among these 60 sources, 51 are found at high latitudes ($|b|>10\degree$), and 38 are detected at $>10\sigma$. 
Additionally, we find that out of the 450 2FLE sources 45 are TeV emitters, and 6 are spatially coincident with a 1LHAASO source.
The 2FLE helps us pave the way for reducing the ``MeV gap", and provides the most sensitive catalog between 20-200 MeV to date.

\end{abstract}


\section{Introduction}\label{sec:intro}
Space-borne gamma-ray astronomy began with the detection of cosmic gamma rays at $E>50$\, MeV by \textit{Explorer 11} after being launched in 1961 (\citealp{kraushaar1962searchexplorer11,kraushaar1965explorer11}). Over the 60 years since then, gamma-ray detections in the soft, MeV energy regime ($E\sim 0.1-100$ MeV) have continued to advance. Shortly after \textit{Explorer 11}, gamma-ray bursts were detected at $E \sim 1$\, MeV energies by using a constellation of \textit{Vela} satellites (\citealp{Vela,grb_vela}). The MeV sky continued to be explored in the 1970s and 1980s by \textit{OSO-3} ($E>50$\,MeV; \citealp{oso-3}), \textit{SAS-B} ($E=20 \, \text{MeV}-1\, \text{GeV}$) aboard \textit{SAS-2} (\citealp{SAS-B}), and \textit{Cos-B} ($E=2\,\text{keV}-5\,\text{GeV}$; \citealp{COS-B,cosb_large}). Observations of the MeV sky continued to be expanded by the launch of the \textit{SIGMA} Telescope ($E = 35\,\text{keV}-1.3\,\text{MeV}$) whose mission lasted from 1989-1997 (\citealp{sigma_results,sigma_death}). Launched in 1991, the Imaging Compton Telescope (COMPTEL) 
aboard the Compton Gamma-Ray Observatory (CGRO) could observe the 1 $-$ 30 MeV sky and detected 32 steady state sources for the first time in gamma rays (\citealp{comptel,Comptel_catalog}). 
CGRO was followed by the 21st century missions of \textit{INTEGRAL} ($E = 15\,\text{keV}-10\,\text{MeV}$; \citealp{INTEGRAL}), \textit{AGILE} ($E = 30\,\text{MeV}-50\,\text{GeV}$; \citealp{AGILE}) , 
 and \textit{Fermi} with both the Large Area Telescope (LAT, \citealp{Fermi_LAT_atwood}) and the Gamma-ray Burst Monitor (GBM, \citealp{fermi_gbm}).

Even with a rich historical record, the sensitivity limits in the MeV regime have left the MeV sky largely underexplored compared to X-ray, GeV, and very-high-energy (VHE) gamma-ray regimes. Three major reasons for this are: 1) the dominant form of photon-matter interaction in this energy regime is Compton scattering, which makes determining the direction of the original photon challenging as well as restricting the effective area of instruments (\citealp{ASTROMEV,AMEGOX}); 2) the instrumental background is very high due to the activation of irradiated materials in space, which limits instrument sensitivity; and 3) technological constraints of silicon-pixel sensors have only recently started to be overcome
\citep{thompson2015spacedetectorsgammarays,brewer2021astropixinvestigatingpotentialsilicon,Brewer_2021}. 

Currently, the most sensitive instrument in the MeV regime is \textit{Fermi}-LAT (\citealp{Fermi_LAT_atwood}), which covers an energy range from 20 MeV $-$ 2 TeV. The \textit{Fermi}-LAT low energy catalog (1FLE; \citealp{1fle_principe}) was the first \textit{Fermi} catalog to report sources detected from $E=31.6\,\rm MeV$ to $E=100\,\rm MeV$. Utilizing the background-independent wavelet transformation code, PGWave (\citealp{pgwave}), the 1FLE detected 198 sources with a significance of $>3\sigma$ using 8.75 years of \textit{Fermi}-LAT data (\citealp{1fle_principe}). 

Previous general LAT catalogs, such as the \textit{Fermi} Large Area Telescope Fourth Source Catalog \citep[4FGL,][]{4fgl}, are more representative of the GeV sky rather than the MeV one because the energy-dependent LAT point-spread function (PSF) and the effective area are optimal near 1 GeV. Yet, \textit{Fermi}-LAT extends down to 20 MeV and now has more than 14 years of observations. 
Hence, dedicated MeV \textit{Fermi}-LAT catalogs are necessary to accurately study MeV-bright astrophysical objects.

A vast number of important astrophysical phenomena are expected to be present in the MeV band (\citealp{COSI,ASTROMEV,AMEGOX}).  
{To date, we know of }hundreds of both Galactic and extra-galactic sources which show a rising hard X-ray spectrum (photon index, $\Gamma_{\rm X}<1.5$) and falling $\gamma$-ray one ($\Gamma_{\rm \gamma}>2$) (\citealp{Gotthelf_2017,Ghisellini_2019,4fgl}). The majority ($>$80\,\%) of the 5835 sources reported in the 4FGL-DR4 catalog (\citealp{4fgl_dr3,4fgl-dr4}) have a power-law spectral index $>$2. It is thus clear that most LAT-detected $\gamma$-ray sources emit most of their energy in the MeV rather than the GeV band (\citealp{Gotthelf_2017,Ghisellini_2019,4fgl}).

As for Galactic sources, there are three primary classes that are expected to be present in the MeV band: pulsars (PSRs), pulsar wind nebulae (PWN), and supernova remnants (SNRs) (\citealp{fryer2019catching,harding2019prospectspulsarstudiesmev}). Young rotation powered pulsars (RPPs) have been detected at hard X-ray energies. These young pulsars exhibit non-thermal emission with hard spectra, which strongly indicates that their peak spectral power is in the MeV range (\citealp{harding2019prospectspulsarstudiesmev}). These observations match current theories which propose that the polar caps (or other regions) of young pulsars are the environments which cause electron-positron pair cascades which produce synchrotron radiation in the outer magnetosphere. The highest energies the synchrotron radiation is anticipated to peak at is a few MeV (\citealp{harding2015synchrotron_mev_pulsar_theory}). Additionally, middle-aged pulsars can host synchrotron halos whose escaped electrons and positrons inverse Compton scatter (IC) off of the interstellar radiation field (ISRF) which are observable in the MeV band (\citealp{dimauro2019prospects}), making them great objects to constrain the complex physics of pulsar environments. Moreover, the relativistic pulsar wind can power  PWNe (\citealt{amato2020theorypulsarwindnebulae}), whose synchrotron radiation can extend to MeV energies, such as seen for the famous Crab PWN.  As for SNR, there is much interest in the MeV regime. This is due to their environments being potential sites of accelerators of cosmic ray protons (\citealp{AMEGOX}). The interaction of protons with interstellar material (such as gas and dust) produces neutral pion decay $(\pi^0 \rightarrow 2\gamma)$ with each photon having an energy of 67.5 MeV in the $\pi^0$ rest frame.

Beyond our Galaxy, accreting supermassive black holes known as active galactic nuclei (AGN), are predicted to be the dominant class of MeV detected sources.
Blazars are a special type of AGN with a relativistic jet pointing along our line-of-sight, usually identified by their Doppler-boosted gamma-ray emission (\citealp{urry_padovani}).
Blazars have a characteristic two-hump spectrum, where the lower energy hump (radio to X-rays) originates from synchrotron radiation of electrons, while the higher energy hump (X-rays to gamma rays) originates from either inverse-Compton scattering of photons off of electrons or hadronic processes such as pion decay (\citealp{ghisellini1989bulk_ssc,Bottcher_2013_lep_had_model,Murase_2016_hidden,costamante2019blazars}). Based on previous gamma-ray and X-ray surveys, we anticipate that a MeV-dedicated catalog will contain a large number of blazars whose inverse-Compton emission is peaked in the MeV, primarily flat-spectrum radio quasars (FSRQs). These MeV blazars are the most powerful of the blazar class and can be found when the Universe was barely 1-2 billion years old (\citealp{ajello2009evolution,Ghisellini_2010,lea_highz_blazar,lea_bass}).
The MeV band is also of interest because one may detect extreme BL Lacs, whose synchrotron peak is in the hard X-ray/MeV band (\citealp{costamante2001extreme_bll}).

Furthermore, neutrinos may plausibly be produced from hadronic processes in the corona of the AGN, and such mechanisms may lead to a high flux of MeV photons, resulting from electromagnetic cascades \citep{Murase_2020,Kurahashi_2022,Murase_2022,ajello2023disentangling}. Thus, MeV sources may be indicative of neutrino sites. The detection of MeV-bright sources could help in the investigation of neutrino sources, and advancing the relatively new field of multimessenger astronomy, where different ``messengers" (e.g. photons, cosmic rays, neutrinos, gravitational waves) are used in tandem to identify and characterize astrophysical sources \citep[\hspace{-0.2cm}][]{icecube2018neutrino,icecube2018multimessenger,aartsen2020time,icecube2022evidence,cutini2023fermi}.

In this paper, we present the Second Catalog of Low-Energy \textit{Fermi}-LAT sources (2FLE). 
The manuscript is structured as follows: Sections \ref{sec:intro}, \ref{sec:data_analysis}, and \ref{sec:pipeline} review our data selection, analysis method, and a few important systematic tests that were performed.  Section \ref{sec:source_assoc}  discusses the method utilized for determining associations to 2FLE sources. Section \ref{sec:2fle_descrip} describes the catalog and its major results while Section \ref{sec:summary} summarizes the main takeaways of this work.

\begin{figure*}[t]
    \centering
    \includegraphics[width=1\textwidth]{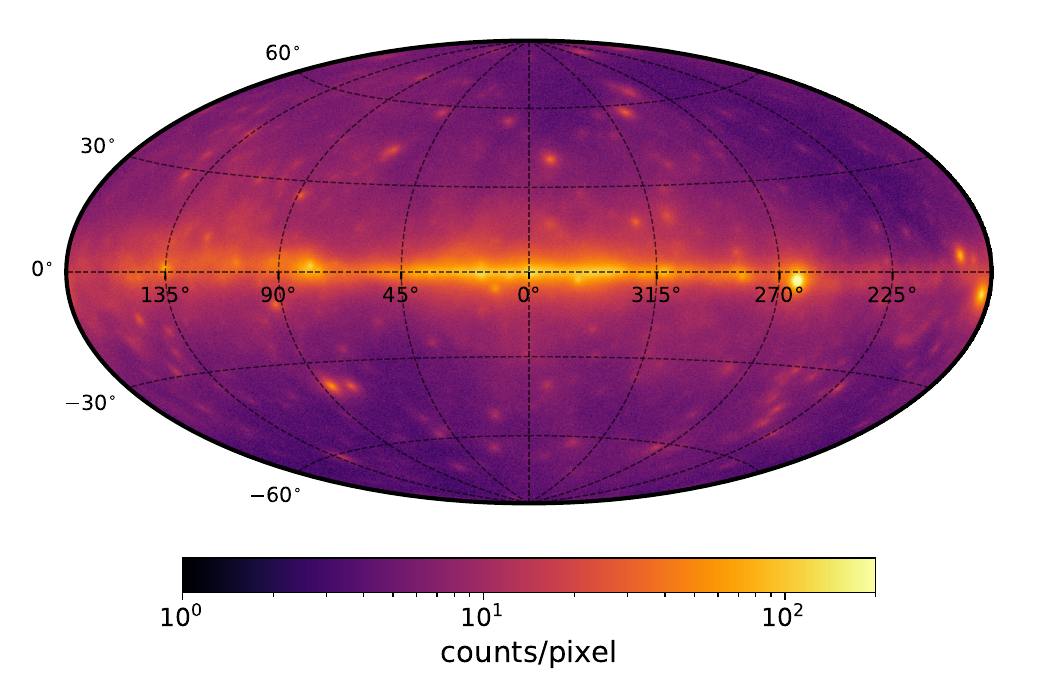}
    \caption{An all-sky Hammer-Aitoff projection in Galactic coordinates of all PSF3 events detected by \textit{Fermi}-LAT over 14 years at $E=20-200$ MeV. The map has been produced with the same data selection as in Section~\ref{sec:data_selsection} with a $0.1\degree$ pixel size.}
    \label{fig:cmap_20_200MEV}
\end{figure*}

\section{Data Analysis} \label{sec:data_analysis}

\subsection{Data Selection and Software}\label{sec:data_selsection}
 The data for this catalog includes source class photons from 20$-$200 MeV detected during the first 14 years of the \fermi-LAT mission (2008 August 04 15:43:36 UTC to 2022 August 02 21:53:15 UTC).
The minimum energy is set to 20~MeV to benefit from the significant amount of effective area\footnote{See the Fermi-LAT performance page \url{https://s3df.slac.stanford.edu/data/fermi/groups/canda/lat_Performance.htm}} below the 1FLE threshold (30~MeV).
 The maximum energy is selected so that we can cover one decade of energy. As a consequence, we benefit from a narrower PSF above 100 MeV.
 For the majority of this time interval, \fermi \ was operating in an all-sky survey mode. Interruptions included passing through the South Atlantic Anomaly (SAA), poor data quality, solar flares, bright GRBs, and a suspension of data acquisition due to a variety of anomalies\footnote{See \url{https://fermi.gsfc.nasa.gov/ssc/library/history/}}, including the failure of a solar panel rotation drive. We excise these events from our catalog's data set, leaving $\sim 10.3$ years of analyzed data. An overview of the LAT's performance after 10 years of observations is presented by \citet{Ajello_2021_lat_10year_performance}. 
 
 For our analysis we use Pass 8 \citep[P8R3, the most up-to-date LAT data set available,][]{atwood2013pass,bruel2018fermilatimprovedpass8event} 
 with the best quality of reconstructed photon direction event type (PSF3). The primary limitation in this analysis is the poor point spread function (PSF) at low energies, hence, we select the best event type available. 
 The 95\% containment radius for the PSF3 event type is $\sim20\degree$ for $E=20$\,MeV, $12\degree$ for $E=50$\,MeV, $7\degree$ for $E=100$\,MeV, and $4\degree$ for 200 MeV \footnote{\url{https://www.slac.stanford.edu/exp/glast/groups/canda/archive/pass8r3v3/lat_Performance_files/gPsfAve95Energy_P8R3_SOURCE_V3psf.png}}.
 Figure \ref{fig:cmap_20_200MEV} displays an all-sky counts map of photon events used in this analysis. The Instrument Response Function (IRF) used is \texttt{P8R3\_SOURCE\_V3}.

 \fermi -LAT's energy resolution is reported to be $\sim20\%$ at 100 MeV and $\sim28\%$ at 30 MeV. 
 We set the energy dispersion parameter, \texttt{edisp\_bins=-1}, except for the extragalactic isotropic diffuse emission model\footnote{\url{https://fermi.gsfc.nasa.gov/ssc/data/analysis/documentation/Pass8_edisp_usage.html}}. We tested different values of \texttt{edisp\_bins} in the interval \texttt{[-3,3]} for our analysis, and found that \texttt{edisp\_bins=-1} is sufficient and only introduces a 3\% systematic on the spectral counts near 200 MeV. The binned maximum likelihood analysis deployed in this investigation implemented square pixels with a bin size of $0.1\degree$ throughout.  
 In energy, we adopt 8 bins per decade. The bin centers are reported in Table \ref{tab:psf_energy_percents}.  
 
\begin{figure}[b]
    \centering
    \includegraphics[width=\linewidth]{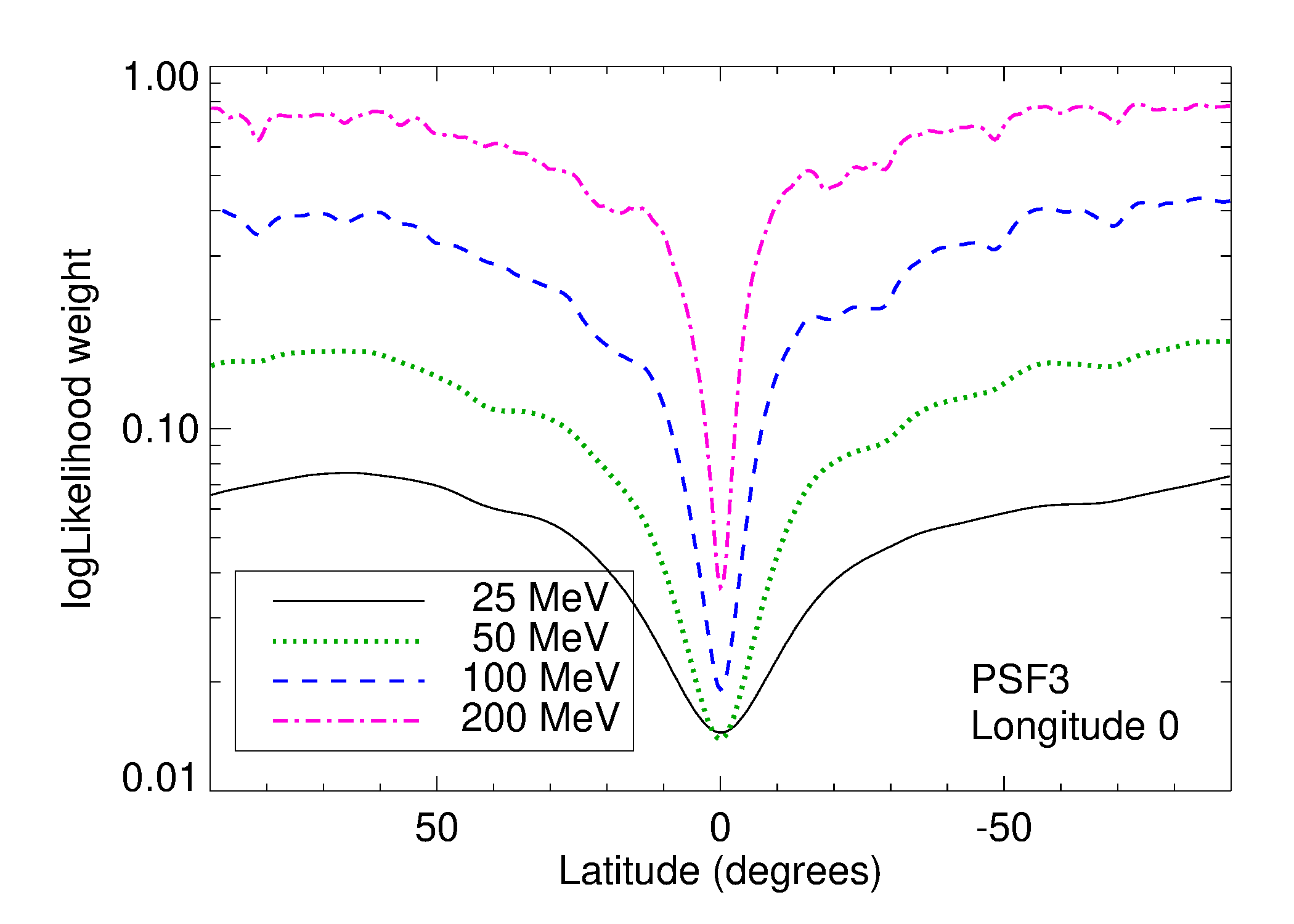}
    \caption{Log-likelihood weights as a function of latitude at 25 MeV, 50 MeV, 100 MeV, and 200 MeV (the Galactic Center is at $0\degree$ latitude).}
    \label{fig:weights}
\end{figure}

 Moreover, since \textit{Fermi}-LAT's low energy regime has uncertainties dominated by systematics, we implement a weighted likelihood analysis\footnote{\url{https://fermi.gsfc.nasa.gov/ssc/data/analysis/scitools/weighted_like.pdf}}, similar to the 4FGL (setting the level of systematic error $\epsilon=3\%$; see Section 3.2 and Appendix B of \citealp{4fgl}), which reduces the statistical influence of low-energy events ($E<100$\,MeV) 
 hence accounting for systematics both on the Galactic diffuse model and on the sources' PSF calibration and imperfect spectral representation. This is especially relevant in regions such as the Galactic Ridge. For all energies $E<50$\, MeV, the weight is $<0.2$ for all locations in the sky, while at $E\sim200$\,MeV, the weights can reach values $>0.8$ (see Figure~\ref{fig:weights}). 
  To limit contamination from gamma rays produced from cosmic rays interacting with the Earth's atmosphere, we implement a zenith angle cut of 80$\degree$. 

For this analysis we employ  \textit{pyLikelihood} and \textit{fermipy} (v0.19.0; \citealt{fermipy_package}) based on \textit{fermitools} (v1.2.23) which are standard LAT Science Tools.

\renewcommand{\arraystretch}{1} 
\setlength{\tabcolsep}{0.95pt}  
\begin{deluxetable}{cc}[t]
\centering
\tablecaption{Energy bin centers and contribution to a $\Gamma=2$ simulated source}
\tablehead{ \colhead{Central Energy (MeV)}  & \colhead{Percent Contribution to Total Integrated Counts} } 
\startdata
23 & $3.7\%$ \\
31 & $7.5\%$\\
42 & $11.5\%$ \\
55 & $14.2\%$ \\
74 & $15.8\%$ \\
98 & $16.3\%$ \\
131 & $16.1\%$ \\
175 & $14.9\%$ \\
\enddata
\end{deluxetable}\label{tab:psf_energy_percents}

 \subsection{Diffuse, Solar, and Lunar Templates}\label{SunandMoon}

 We include the same Galactic diffuse (\texttt{gll\_iem\_v07.fits}) and extragalactic isotropic diffuse emission (\texttt{iso\_P8R3\_SOURCE\_V3\_PSF3\_v1.txt}) templates used in the 4FGL catalog\footnote{Provided by the Fermi Science Support Center here: \url{https://fermi.gsfc.nasa.gov/ssc/data/access/lat/BackgroundModels.html}} (\citealp{4fgl}). 
The isotropic and galactic diffuse models are defined, respectively, down to 30 MeV and 50 MeV and need to be extrapolated to 20 MeV for this analysis. For the Galactic emission we rely on the power-law extrapolation automatically performed by the {\it Fermitools} down to 20 MeV from the first two energy bins (50 and 65 MeV) of the model. 
 For the isotropic model, we introduce a broken power law with a break at 30 MeV,  with an index below 30 MeV that has been optimized for every ROI to reduce the energy residuals. No changes were made to the morphology of the diffuse models.   The only sources we use \textit{a priori} in our catalog are the extended sources reported in the 4FGL Data Release 4 (4FGL-DR4; \citealp{4fgl-dr4}), whose spectral parameters are kept frozen to the 4FGL ones throughout the analysis. 
 
Since they are bright in gamma rays, we model the Sun and Moon (as described in \citealp{sun_moon_limb}). Using \texttt{gtsuntemp}, we model the contribution of these two components 
with two templates: a solar inverse-Compton emission model (pixel size $0.25\degree$), and a solar and lunar disc model (pixel size $0.125\degree$). The first template is solely the inverse Compton emission of solar photons with cosmic-ray positrons and electrons, resulting in an extended halo from the Sun (\citealt{Ng_2016_solar_templates}).
The solar and lunar disc model is representative of the interaction of hadronic cosmic rays with the Sun and Moon. For the Moon, bombardment of cosmic rays on the Moon's surface initiates particle cascades that lead to the decay of secondary neutral pions and kaons into gamma rays (\citealp{moon_lunar_gamma_rays,moon_2016}). For the Sun, the hadronic interactions occur in the Sun's corona \citep{sun_2025}. These phenomena  persist for the Sun and Moon as they move across the sky, strongly emitting MeV-dominant photons. Therefore, regions studied along the ecliptic are susceptible to interference from the presence of these bright gamma-ray emitting bodies.

Both templates were calculated using the same event selections and zenith angle cuts as described in Section \ref{sec:data_selsection}.  The solar contributions are formally defined down to 1 MeV, while the lunar template was extrapolated down to 20 MeV (from 30 MeV). Both have been extended to cover 14 years of \textit{Fermi}-LAT mission time. Both of these components are fixed in the model.

\section{Detection Pipeline} \label{sec:pipeline}
We tessellated the entire celestial sphere in a total of 266 regions of interest (ROIs). We define ROIs as $20\degree\times20\degree$ square regions (AIT projection) that are spaced in intervals of  $15\degree$ in both Galactic latitude ($b$) and longitude ($l$). 
To avoid oversampling the North and South Galactic poles, we cover both regions with a single ROI, centered at ($b=+90\degree$, $l=0\degree$) and ($b=-90\degree$, $l=0\degree$), respectively. The centers of the other ROIs are determined in the following way: starting from the North Galactic pole, we descend in latitude by $15\degree$ and then arrange 24 ROIs whose centers are separated by $15\degree$ in longitude to cover the full $360\degree$ sky circle at that latitude. We proceed this way until we reach the South Galactic pole. The $20\degree\times20\degree$ size of the ROI guarantees that every ROI has at least a $5\degree$ overlap with each of its four adjacent ROIs (1-left, 1-right, 1-up, 1-down). This layout also ensures that
the Galactic center ($b=0\degree, l=0\degree$) is centered in a single ROI. 
This ROI configuration has been chosen to allow for a proper characterization of sources that are near the edges of an ROI. Various ROI sizes were tested in a few different configurations, including different sizes for different regions of the sky. The final ROI distribution and size was selected to center ROIs on the Galactic plane, account for larger PSF at lower MeV energies and faint sources, while ensuring all ROI fits would properly converge. Much larger ROIs cannot be used because the fits become unstable. Additionally, we set the maximum distance from the ROI center to $30\degree$ (in \textit{fermipy} \texttt{src\_roiwidth=$30\degree$}) to account for the influence of sources in the surrounding regions.

Since the sky is tessellated into numerous ROIs, final fits of the isotropic and Galactic diffuse model parameters are evaluated for model agreement with adjacent ROIs. ROI diffuse index parameters are considered in disagreement if their reported values differ by $\geq3\sigma$. The overwhelming majority of ROIs (261/266) have both isotropic and Galactic diffuse index agreement with at least 2 adjacent ROIs. Moreover, 85\%-92\% of neighboring ROIs have normalization of both the isotropic and Galactic diffuse emission compatible with each other.

The spectrum of each source is modeled by using a power law as described by 
\begin{equation}
    \frac{dN}{dE} = N_0 \left(\frac{E}{E_0}\right)^{-\Gamma}
\end{equation}

where $N_0$ is the prefactor (also known as normalization), $E$ is energy in MeV, $\Gamma$ is the index, and $E_0$ is a reference energy kept fixed at 1 GeV. The final catalog reports the energy at which uncertainty on differential flux is minimal (Pivot\_Energy, see Table~\ref{tab:cat_col_description}), a value optimized a posteriori after the best-fit spectral parameters are computed for all sources. 

For every ROI, the point source detection pipeline is as follows:
\begin{enumerate}
    \item Using the configuration listed in Section~\ref{sec:data_selsection}, a Test-Statistic\footnote{The TS is defined as $-2\log(\rm L_0/L)$, where $\rm L_0$ is the likelihood for the null hypothesis, and $\rm L$ is the likelihood for the alternative hypothesis.} ($TS$) map is generated to first search for bright, significantly detected ($TS>$64; $\gtrsim8\sigma$) sources. To generate the $TS$ map, we assume that the spectral shape of the underlying sources is a power law with spectral index value, $\Gamma=2$. The only free parameters at this stage are the isotropic and Galactic diffuse components.
    
    \item Once found, the sources are localized (the localization and positional uncertainty determination are detailed in Section~\ref{sec:loc}), and their spectral parameters are optimized by fitting the sources individually (i.e.~freezing the parameters of all the other sources in the ROI besides the diffuse components). When the fit has converged successfully, the new sources are inserted into the ROI model, unless they are detected with a significance of $TS<25$, in which case they are deleted from the model. This is to avoid a lower significance source from affecting the localization of a more significant source detection.
    
    \item The same procedure as 1 and 2 is followed iteratively to localize and add sources into the model detected at an intermediate significance ($TS>$36) as well as a lower significance ($TS>$16). At the last iteration (finding sources at $TS>$16), we do not \textit{remove} sources with $TS<25$. We keep all sources (including $TS<16$) until Step 9.

    \item Once all the sources have been added to the ROI model, a global fit is performed. For this fit (using \textit{fermipy}'s \texttt{fit()} method),  the normalizations of all sources (except for extended sources, as well as the Sun and Moon) are kept free. 
    
    \item Now that we have some knowledge of how the sources are distributed in the ROI, we employ a relocalization method to find their best-fit positions.
    Starting with the most significant sources, going one by one from the highest to the lowest $TS$, we: (a) remove the source from the ROI model; (b) run the source finding method again (detailed in 1) to refine the localization using the best-fit spectral index found for the source in step 2; (c) optimize the spectral parameter (as done in 2). After every source is individually fit, a global fit is run, where the normalization of all sources and the Galactic and isotropic diffuse models are free to vary in the fit. We note that ideally the spectral indices of all sources in the ROI should be also kept free in this fit. However, due to the complexity of our ROIs, this is computationally unfeasible. We therefore caution the catalog users on the fact that spectral parameters uncertainties may be slightly underestimated (by a factor of $\sim20\%$) for sources with close separations ($<3\degree$, about $16\%$ of sources in the catalog).
    
    \item If a source's spectral parameters do not converge during the re-localization step, then we iterate through different starting spectral index values in the range $\Gamma=[0.5-4.5]$, in steps of 0.5. If the source still does not converge, it is kept with a spectral index set to $\Gamma=2$. 
    
    \item After re-localizing all originally detected sources, and running a global fit, for any source that failed to converge we use the method done in 6.

    \item The normalization of all sources and the Galactic and isotropic diffuse components are finally optimized at once in a global fit.
    \item All sources located at $<2.5\degree$ from the edge of the ROI are removed from the model. We also remove any source with a likelihood of $TS<25$. 
    \item The individual lists of remaining sources are then merged together into one catalog
    \item Duplicate detections of the same source are removed establishing the final catalog. For details on duplicate detections see Section \ref{sec:duplicate}.
\end{enumerate}

\subsection{Effective PSF of point source with $\Gamma=2.0$} \label{sec:effective_psf}
To characterize the LAT PSF at these energies, we simulated a bright source with a spectral power law index of $\Gamma=2.0$ (a typical value for most extragalactic sources) in the center of an arbitrary extragalactic ROI. No diffuse or point-source background sources were included.

\begin{figure}[!tbp]
    \centering
    \subfloat[]{
    \includegraphics[width=\columnwidth]{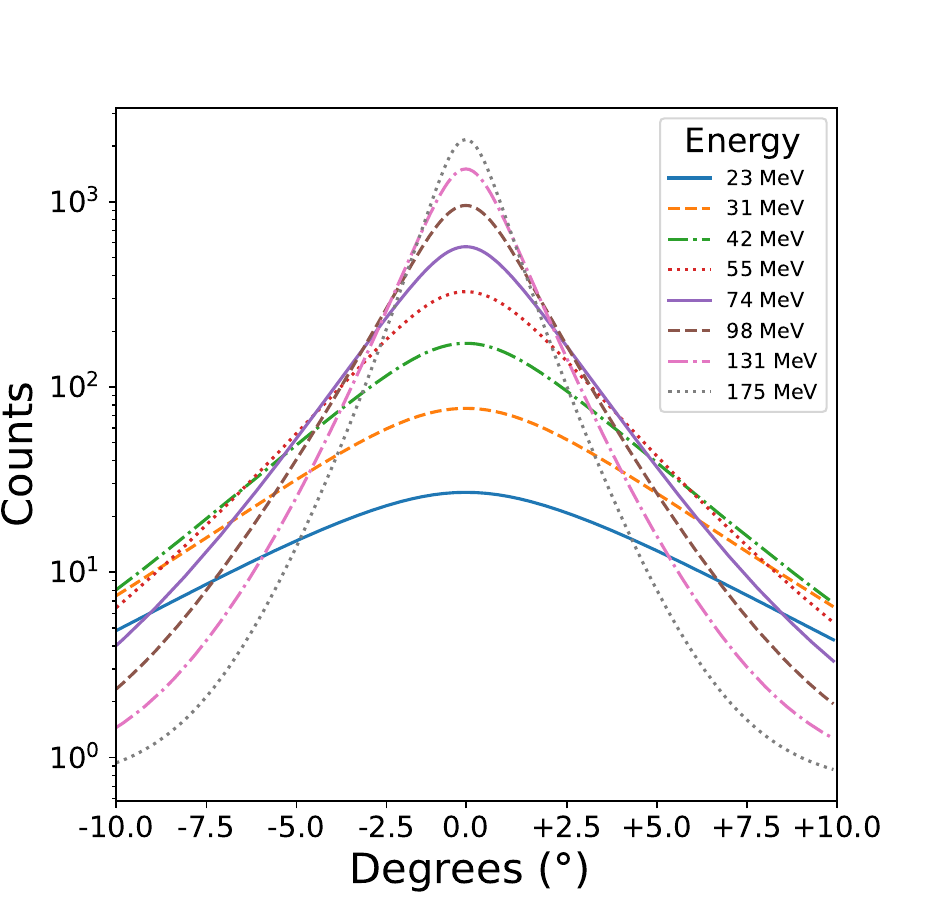}\label{fig:individual_psf}}
    \hfill
    \subfloat[]{
    \includegraphics[width=\columnwidth]{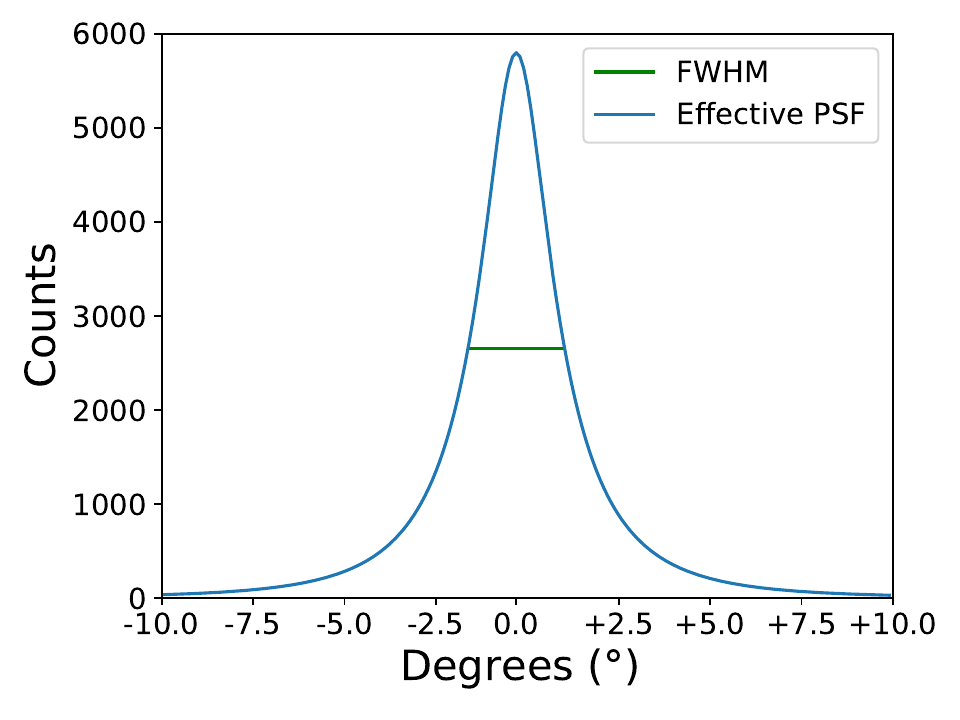}\label{fig:total_psf}}
    \caption{\textbf{Top (a):} 1D slices of the PSF as a function of energy for a simulated source with $\Gamma=2.0$. The y-axis is in log scale. 
    \textbf{Bottom (b): } Sum of all the slices above representing the effective PSF for a source with $\Gamma=2.0$. The green line represents the FWHM calculated for the effective PSF. The y-axis is in linear scale.}
\end{figure}

Figure \ref{fig:individual_psf} shows each 1D PSF for different energy bins, while Figure \ref{fig:total_psf} sums them together, resulting in the effective PSF of the instrument at these energies for a source with $\Gamma=2.0$. We calculate the full-width-half-max (FWHM) of the effective PSF to be 2.4$\degree$, seen as the green line in Figure \ref{fig:total_psf}. 
Table \ref{tab:psf_energy_percents} reports the fraction of counts contributed in 8 energy bins over the 20-200 MeV band.

\begin{figure*}
    \centering
    \includegraphics[width=1.2\columnwidth]{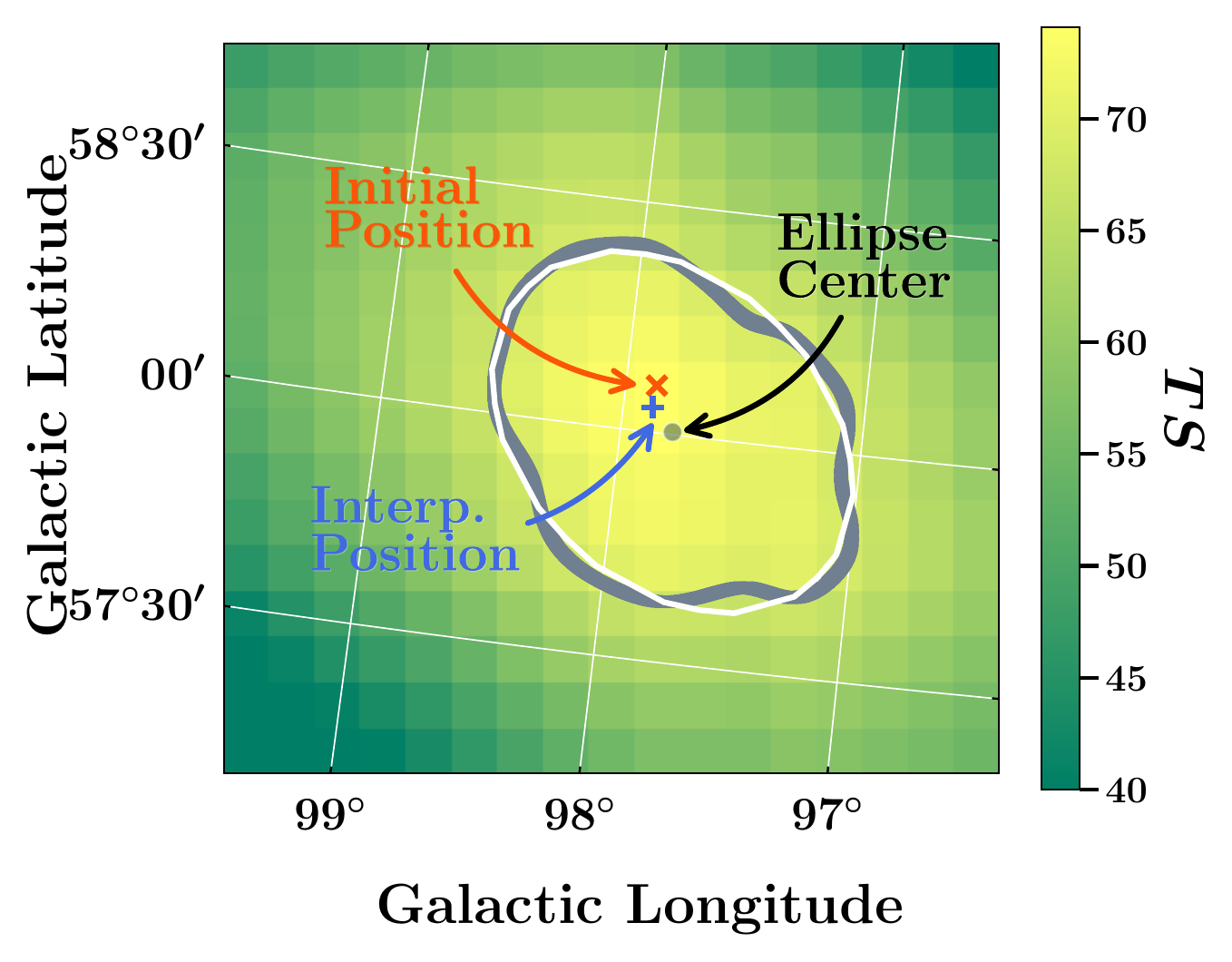}
    \caption{Example of the source localization and positional uncertainty region determination using the $\Delta TS$ interpolation method (see Section~\ref{sec:loc}). The orange cross indicates the initial position of the source identified by \textit{fermipy} source finding algorithm. The map shown in the background is the $TS$ map around the location of the source (the maximum $TS$ of the source in this map, hence of the color scale, is 74). The blue plus sign marks the new source position found by interpolating the $\Delta TS$ map around the peak and defining a subgrid of points finer than the ROI pixel size to locate the true minimum of $\Delta TS$ (i.e.~true $TS$ maximum). To find the 95\% positional uncertainty, we select all the points from the true minimum for which $\Delta TS=5.99$ (gray jagged line) and fit them with an ellipse (white line). The center of the ellipse is shown as a gray dot.}
    \label{fig:tess}
\end{figure*}

\subsection{Position \& positional uncertainty}\label{sec:loc}
Since the LAT PSF at the energies covered by our catalog is large and  non-Gaussian, to determine the sources’ best-fit position we developed a technique that relies on interpolation of the TS maps. At first, following the standard \textit{fermipy} methodology, the TS map in every ROI is scanned for significant peaks – i.e. points in the map at $TS>TS_{\rm thresh}$ for $TS_{\rm thresh} = 64, 25, \text{and } 16$. 
For every peak found, we construct the $\Delta TS$ map:
\begin{equation}
    \Delta TS = TS_{\rm peak, i}-TS_{j},
\end{equation}
for all the $j$ pixels around the $i_{th}$ peak. We then interpolate\footnote{We rely on the {\it scipy} bivariate spline interpolation.} the $\Delta TS$ values around the peak to map out the $\Delta TS$ distribution. Next, we define a subgrid of points finer than the original ROI pixel size and use the interpolation to find the location of the true minimum of the $\Delta TS$ map (or true maximum of the $TS$ map) around that point. This gives us the best-fit coordinates for the source. 

Once the best-fit location is found, we need to define an error region. 
Therefore, we recalibrate our interpolation grid based on the value of the true $\Delta TS$ minimum to find the region of points at a $\Delta TS=5.99$ (corresponding to 95\% confinement radius for 2 d.o.f.)
around each best-fit location. Finally, we fit this region with an ellipse to determine our 95\% positional uncertainty. 

This uncertainty region is often elliptical (rather than circular). Hence in the catalog we report the semi-major and semi-minor axis of the ellipse corresponding to the 95\% positional uncertainty. A visualization of the interpolation is shown in Figure \ref{fig:tess}.

\subsection{Duplicate Detections}\label{sec:duplicate}
Due to the ROIs overlapping on the sky, a detected source in one ROI may also be detected in an adjacent ROI. Sources on the edge of two ROIs naturally will be detected in more than one ROI, with one case closer to the center of one of the adjacent ROIs. In order to avoid reporting duplicate detections we implement a simple duplicate finding code based on spatial and spectral coincidence. Essentially, if two or more sources are spatially coincident and originate from different ROIs, they are considered potential duplicate candidates. For each set of overlapping sources from separate ROIs (which can sometimes be $>2$), we compare their spectral indices to one another. If their spectral indices agree within 2$\sigma$ they are considered a match and the one with the highest $TS$ is kept. The 2$\sigma$ threshold was selected based on the simulations discussed in Section \ref{sec:sim_r95}. 
The remaining spatially coincident sources that were not spectrally coincident typically contain one of the sources having convergence issues and reporting a $\Gamma\sim0$. Even in these cases, we keep the source with the highest $TS$, which typically has better constrained parameters. We do not find cases of well-converged, spatially coincident sources with drastically different spectral properties.

\subsection{95\% Uncertainty Region Calibration} \label{sec:sim_r95}
\begin{figure}
    \centering
    \includegraphics[width=\columnwidth]{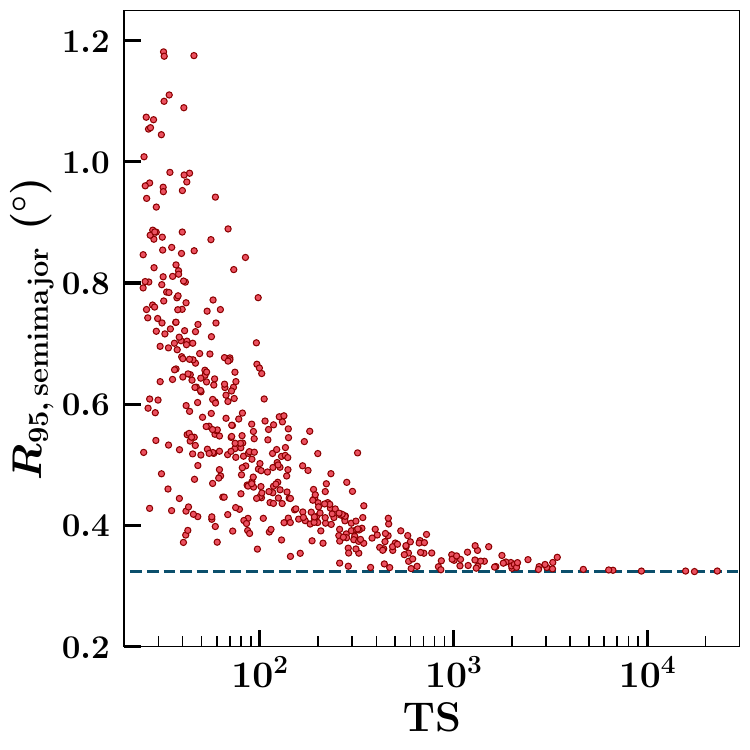}
    \caption{Distribution of \texttt{Conf\_95\_SemiMajor} ($R_{95,\,\rm semimajor}$) corrected by systematic errors (Section~\ref{sec:sim_r95}) as function of sources' TS values for all 2FLE sources. The blue dashed line denotes the minimum value of $R_{95,\,\rm semimajor}=0.32^{\circ}$.}
    \label{fig:tsr95}
\end{figure}

As done in the 4FGL, as well as other investigations such as \citet{joffre2024historical_arxiv}, the 95\% uncertainty region radius (\rad) needs to be calibrated. For this discussion, the original, uncalibrated radius was determined by the method described in Section \ref{sec:loc}. The initial 95\% semi-major axis uncertainty region radius will be denoted by $R^{i}_{95}$ while the corrected value which we report in the catalog is reported as \rad. We simulated 10 realizations of the 20$-$200 MeV sky. Utilizing the \textit{fermipy} library method \texttt{gta.simulate\_roi}, we simulated each ROI by generating an array of Poisson random numbers with expectation values pulled from the model cube of the instance of the binned analysis. We utilized the same template files in our catalog as described in Section \ref{sec:data_analysis}, adding all the 4FGL-DR4 sources to the initial configuration file with their best-fit spectrum extrapolated to this energy range. Next we ran our detection pipeline (described in Section \ref{sec:pipeline}) to detect the simulated sources. Since our simulation only uses 4FGL-DR4 data, every source identified by our pipeline is expected to capture a simulated 4FGL-DR4 source. 
We assume that the closest simulated 4FGL-DR4 source to each detection was the source that was being detected by our method, unless the ratio of angular distance/$R^i_{68}$ was $>3$. The selected 4FGL source  
was verified as the proper corresponding detected simulated source by comparing the extrapolated $20-200$\,MeV flux from the 4FGL-DR4 to what was detected such that the flux ratio is $\sim1$. 
We identify two correction factors\footnote{The initial assumption of a 2D offset follows a normal law. Converted to polar coordinates this can be generalized to a Rayleigh distribution. The 95\% containment radius ($R_{95}$)  is $-2\text{ln}(0.05)\sigma$} $m$ and $n$ where $m$ is the 
multiplicative correction factor for 
the $R^{i}_{95}$ value, while $n$ corrects for a systematic offset such that

\begin{equation}\label{eq:r95}
    R_{95} = \sqrt{(R^{i}_{95}*m)^2 + n^2}
\end{equation}
Using this adjusted value combined with the angular separation, $\Theta$, between the 4FGL-DR4 source and our simulated detection, we establish the relation between $\Theta$ and $R_{95}$:
\begin{equation}\label{eqn:y}
    y = \frac{1}{2} \left( \frac{\Theta}{R_{95}/2.45}\right)^2,
\end{equation}

where 2.45 is the conversion factor from $R_{95}$ into the equivalent 1D Gaussian standard deviation \citep[$\sigma$, e.g.,][]{Andrae_2010}. 
Using this relation, 
we compare the histogram of observed $e^{-y}$ values to the (theoretical) exponential probability distribution $p(y)=e^{-y}$ \citep[e.g.][]{Czesla_2023},
adjusting $m$ and $n$ until the $\chi^2$ is minimized. 
This follows a similar method done by \citet{joffre2024historical_arxiv}. The best-fit values are of $m=1.20$, $n=0.3\degree$. These corrections to systematic errors in the 95\% uncertainty measurement are applied to all of the originally found $R^i_{95}$ values such that our catalog only reports the corrected values in the columns \texttt{Conf\_95\_SemiMajor} and \texttt{Conf\_95\_SemiMinor} so all values are $>0.3\degree$. Figure~\ref{fig:tsr95} shows the distribution of corrected \texttt{Conf\_95\_SemiMajor} as function of sources' TS values.

\subsection{Simulations with Alternative Galactic Diffuse Models} \label{sec:altgal_diffuse_sim}
Based on concerns of mistaking interstellar diffuse emission for a source, we additionally tested simulations using 3 alternative Galactic diffuse models which are different from what we used for the 2FLE. The alternative models used were the previous version of the current Galactic diffuse model (\texttt{gll\_iem\_v06}) and 2 alternative models developed for the First \textit{Fermi}-LAT Supernova Remnant Catalog (\citealp{acero2016firstSNR}; \textit{Lorimer\_z10\_Ts100000\_v5\_mapcube\_total\_all\_P8.fits} and \textit{SNR\_z4\_Ts150\_v5\_mapcube\_total\_all\_P8.fits}). Of these three models, we refer to them as \texttt{gll\_v06}, \texttt{Lorimer}, and \texttt{SNR\_z4}, respectively, for the rest of the paper. The description of how \texttt{Lorimer}, and \texttt{SNR\_z4} were made can be found in Section 2.4.2 of \citet{acero2016firstSNR}. We selected the two models that had the most different parameters from one another, but were representative of each type of general model (\texttt{Lorimer} versus \texttt{SNR}) that was developed in \citet{acero2016firstSNR}.

\renewcommand{\arraystretch}{1} 
\setlength{\tabcolsep}{5pt}    
\begin{deluxetable}{cccc}[t]
\centering
   \tablecaption{Alternative Diffuse Model Spurious Source Detection}
    \tablehead{
    \colhead{Model}  & \colhead{Spurious Detections}  & \colhead{$|b|>35\degree$} & \colhead{$|b|<35 \degree$} }   
\startdata
\toprule
\texttt{gll\_v06} & 20 (4.4\%) & 6 (2.6\%) & 14 (6.3\%)\\
\texttt{SNR\_z4} & 91 (20.2\%) & 47 (20.5\%) & 44 (20.0\%)\\
\texttt{Lorimer} & 67 (14.9\%) & 21 (9.2\%) & 46 (20.8\%)\\
\midrule
Average & 13\% & 11\% &  16.4\% \\
\bottomrule
\enddata
\footnotesize
\tablecomments{Each model tested is listed, with their combined results reported in the bottom `Combined' row. The number reported is the number of spurious sources detected with the alternative diffuse model. The percentage depicted in parentheses is what fraction the number of spurious detections make up of the number of sources reported in the 2FLE at those latitudes. 
The last row is the average of the percentages of spurious sources for the 3 models.}
\end{deluxetable}\label{tab:alt_model_results}

We tested all ROIs to evaluate spurious source detection across the entire sky. Again, we only populate the simulated regions with the Galactic and isotropic diffuse emission as well as the Sun and Moon. Simulating the region using one of the alternative models with \texttt{gta.simulate\_roi}, we then ran our source detection pipeline using the same diffuse model utilized to build the 2FLE (\texttt{gll\_iem\_v07}) with the same cuts in TS and distance from the edge. We find that the \texttt{SNR\_z4} and \texttt{Lorimer} models produced simulated skies that were the most different from \texttt{gll\_iem\_v07} and had the most spurious sources mistaking interstellar emission for a gamma-ray source. Since Galactic diffuse emission can reach $|b|=35\degree$, we report the number of detections for the entire sky, $|b|>35\degree$, and $|b|<35\degree$ in Table \ref{tab:alt_model_results}. In the Table \ref{tab:alt_model_results} columns, the value in parentheses indicates the percentage associated with the number of spurious sources divided by the number of 2FLE sources detected at those given Galactic latitudes. These percentages reflect the estimated percent of spurious sources reported in the catalog, assuming the true diffuse emission is as different as the models are from \texttt{gll\_iem\_v07}. Since our isotropic and Galactic models are the most current, and were optimized to the data, we anticipate our true spurious detection percentage to be closer to the results from \texttt{gll\_v06}, or $4.4-6.3\%$ (20-28 sources).

Of the 3 separate alternative models tested, the highest TS for a spurious detection was 110, however, more than $60\%$ of all spurious detections were detected at $TS<35$. Therefore, we warn the reader that sources detected with a $TS<35$, especially at $|b|<35\degree$, are most susceptible to being mistaken detections of mismodeled interstellar gas emission. 

Cross-matching the locations of all spurious detections (that remain after distance from the edge and $TS>25$ cuts) from the 3 separate simulations with the 2FLE, 5 sources in the 2FLE were found to be spatially coincident with a spurious source. These sources, are 2FLE J2202.8+5056*, 2FLE J1148.1-3837*, 2FLE J0419.3+3817*, 2FLE J2133.7-0026*, and 2FLE J1113.3-0834*, where asterisk (*) indicates potential confusion with interstellar emission. 
However, we note that 4 of these detections are associated to known gamma-ray sources associated to AGN and not reported as flagged in the 4FGL. The only exception is 2FLE J2133.7-0026, which was detected at $TS=66$ and was found associated to the globular cluster (GLC) M2. This source is located at $b=-35.6\degree$, but due to its significant detection we do not believe it to be a spurious detection. The spurious detection that was found coincident with 2FLE J2133.7-0026 is only detected at $TS=32$.

\subsection{Spectral Energy Distributions}\label{sec:sed_gen}

For every source in the 2FLE catalog we compute the spectral energy distributions using the {\it fermipy} SED tool\footnote{\url{https://fermipy.readthedocs.io/en/latest/advanced/sed.html}}. The SEDs are computed in 3 energy bins: 20$-$43.1 MeV ($E_{\rm mid}=29.3\,\rm MeV$); 43.1$-$92.8 MeV ($E_{\rm mid}=63.2\,\rm MeV$); 92.8$-$200 MeV ($E_{\rm mid}=136.2\,\rm MeV$). For every source, we free the normalization of the background components (\texttt{free\_background=True}) and we fix the spectral index to its best fit over the full interval. 
In the catalog, we report the differential flux in the 3 bins with associated $1\sigma$ uncertainties and the TS values in every bin.

\renewcommand{\arraystretch}{1} 
\setlength{\tabcolsep}{0.9pt}    
\begin{deluxetable}{lc}[t]
\centering
   \tablecaption{2FLE Associations}
    \tablehead{\colhead{Type of Associations} & \colhead{No. of Sources}
    }   
    \startdata
    \hline
    2FLE Sources & 450 \\
    High confidence Associations (ASSOC1) & 254\\
    Low confidence Associations (ASSOC3) & 50\\
    \hline
    Assoc.~with 4FGL-DR4 $\gamma$-ray source (ASSOC\_GAM) & 387\\
    High confidence Associations in 4FGL-DR4 (ASSOC2) & 370\\
    2FLE Uncertainty Encompasses Two 4FGL-DR4 Sources & 53 \\
    \hline
    ASSOC1 with ASSOC\_GAM w/o ASSOC2 & 4\\
    Discrepant Associations between ASSOC1 and ASSOC2 & 13 \\
    ASSOC1 w/o ASSOC\_GAM (See Table \ref{tab:2fle_newblazars})  & 3 \\
    \hline
    2FLE w/o ASSOC1 or ASSOC2  & 73\\
    2FLE w/o ASSOC1, ASSOC2 or ASSOC3  & 29
    \enddata
\end{deluxetable}\label{tab:source_assoc_breakdown}

\renewcommand{\arraystretch}{1} 
\setlength{\tabcolsep}{1pt}    
\begin{deluxetable*}{cccccccccc}[t]
\centering
   \tablecaption{Associated 2FLE sources lacking a 4FGL counterpart listed in order of decreasing TS.}
    \tablehead{\colhead{2FLE Name$^{\dagger}$}  & \colhead{2FLE RA} & \colhead{2FLE DEC.} & \colhead{Conf\_95\_SemiMajor} & \colhead{TS} & \colhead{2FLE Assoc.} & \colhead{Class} & \colhead{RA\_count} & \colhead{DEC\_count} & \colhead{ASSOC\_PROB}\\
    \colhead{}   &  \colhead{($\degree$)}  &  \colhead{($\degree$)}  &  \colhead{($\degree$)}  &  \colhead{}  &  \colhead{}  &  \colhead{}  &  \colhead{($\degree$)}  &  \colhead{($\degree$)}  &  \colhead{}  
    }   
\startdata
\hline
2FLE J0434.8$-$4356 & 68.719 & -43.943 & 0.722 & 73 & PKS 0432-440 & fsrq &   68.514 &	-43.930  & 0.80 \\
2FLE J1105.7+1141 & 166.44 & 11.685  & 0.576 & 70 & GB6 B1102+1149 & bcu   & 166.392&	11.546 &  0.85 \\
2FLE J2133.7$-$0026 & 323.442 & -0.433 & 0.533 &  66 & M 2 & glc &  323.363 &	-0.823   &   0.81 \\
\hline
\enddata
\tablenotetext{$$\dagger$$}{All columns correspond to those reported in the 2FLE (see Table \ref{tab:cat_col_description} for reference).}
\end{deluxetable*}\label{tab:2fle_newblazars}

\section{Source Association}\label{sec:source_assoc}
The association procedure makes use of the Bayesian and likelihood-ratio (LR) methods followed for the FGL general catalogs and described in Appendix B of 1FGL (\citealp{1fgl}) and in the 2LAC paper (\citealp{LAT11_2LAC}), respectively. The multiwavelength catalogs used for the association procedure are listed in the extension table of the catalog's FITS file, described in Appendix~\ref{app:a}. The pipeline does not consider physical distances, but it takes into account the densities of $\gamma$-ray emitting sources on the sky.
The prior parameters needed by the Bayesian method have been kept fixed to the 4FGL-DR4 values and the same counterpart catalogs as in 4FGL-DR4 have been used. The Bayesian method returns 252 high-confidence (association probability $\geq$0.8) associations. The LR method yields 204 associations, and only two of them (for 2FLE J0340.1+3215, ASSOC\_PROB\_LR=0.91, and 2FLE J1458.1-3602, ASSOC\_PROB\_LR=0.91) are missed by the Bayesian method. Therefore, we report a high-confidence association for a total of 254 2FLE sources (ASSOC1 in the catalog's fits file). Since the two methods return almost identical associations, for the 2FLE we only report the probability of association  (ASSOC\_PROB in the catalog's fits file) obtained with the Bayesian method for 252 sources and with the likelihood-ratio method for 2FLE J0340.1+3215 and 2FLE J1458.1-3602.

Given the larger uncertainty ellipses  relative to the FGL catalogs (by a factor often  exceeding 100 in solid angle) inherent in using only low-energy photons instead of the whole set detected by the LAT, the association probabilities derived from the 2FLE  spatial parameters tend to be notably lower  than those derived in the context of 4FGL, affecting the association rate significantly. 
Therefore, we also cross-matched the 2FLE source list with the 4FGL-DR4 to determine how many 2FLE sources have an already known $\gamma$-ray counterpart and, of those, how many have an association in the 4FGL-DR4. In this case, the positional cross-match was performed using the \texttt{Sky Ellipse} tool in the ToPCAT software \citep[v4.9,][]{topcat}. For both the 4FGL and the 2FLE, we use the \texttt{Conf\_95\_SemiMajor}, \texttt{Conf\_95\_SemiMinor} and \texttt{Conf\_95\_PosAng} listed in both catalogs as semimajor, semiminor and angle of the ellipses. This exercise returns 377 matches, of which 326 are unique matches (i.e.~only one 4FGL-DR4 source is found in the 95\% ellipse region), while for 51 2FLE sources we find more than one 4FGL-DR4 source. In the latter case, to select the most probable 4FGL counterpart to the 2FLE source, we compare the 2FLE SEDs (as computed in Section~\ref{sec:sed_gen}) with the 4FGL ones for all the matches and pick the source with the most compatible SED (i.e.~the 2FLE data points are compatible with the 4FGL ones and the spectral parameters are consistent within $1\sigma$). Finally, for the remaining 78 2FLE sources we also search for a 4FGL counterpart within the \texttt{Conf\_95\_SemiMajor} circle and we find 10 more 4FGL matches, 8 of which are unique, while for the other 2 we apply the same SED criteria described above. Overall, 387 2FLE sources have a 4FGL-DR4 counterpart (ASSOC\_GAM in the catalog); of those, 370 have an associated counterpart in the 4FGL-DR4 (ASSOC2 in the catalog). 

This procedure enables us to report 123 extra associations, for a total of 377 2FLE associated sources.  
Given that 53/387 matches with a 4FGL-DR4 source are not unique, this gives us a lower limit of $\sim14\%$ on the percentage of catalog sources suffering from blending, meaning that the sources' localization is impacted by the presence of 2 or more nearby $\gamma$-ray sources. To better quantify this effect, we also crossmatch the MeV bright sources in the 4FGL-DR4 with the 2FLE ones using the same method as described above. From the 4FGL-DR4, we select all the sources that have a $TS_{4FGL,50-300\,\rm MeV}>100$, where $TS_{4FGL,50-300\,\rm MeV}$ is the sum of the squares of the source significance in the first two SED bands listed in the 4FGL catalog \citep{4fgl-dr4}. Crossmatching the 2FLE with these 223 4FGL-DR4 sources returns 179 matches (80\%), implying that blending is affecting $\sim20\%$ of our catalog. 

Some sources show discrepant associations between ASSOC1 and ASSOC2, because the association probability derived with the 2FLE position parameters favors another counterpart (or find one when it was missing) than that found in the 4FGL-DR4 catalog for the corresponding source (reported in ASSOC\_GAM).  Four 2FLE sources (2FLE J0340.1+3215, 2FLE J0746.9+7439, 2FLE J1356.4$-$6143, and 2FLE J2240.3$-$5214) have associations while the corresponding 4FGL-DR4 sources were lacking one. Twelve 2FLE sources are associated with a different AGN than did their 4FGL-DR4 counterparts.  The source 2FLE J2025.5+3335 is associated with a pulsar, whereas its 4FGL-DR4 counterpart was an AGN. 
The association of 2FLE J0235.7$-$3002  is PKS J0236$-$2953,  different from that of 4FGL J0235.6$-$2939, namely 1RXS J023536.7$-$293845 (erroneously reported as PHL 1389 in 4FGL-DR4). 
We also note that there are four 2FLE sources that are associated, but lack a 4FGL counterpart, hence this is the first time they are detected in $\gamma$-rays (see Table~\ref{tab:2fle_newblazars} for details).

\begin{figure*}[t]
    \centering
    \includegraphics[width=\textwidth]{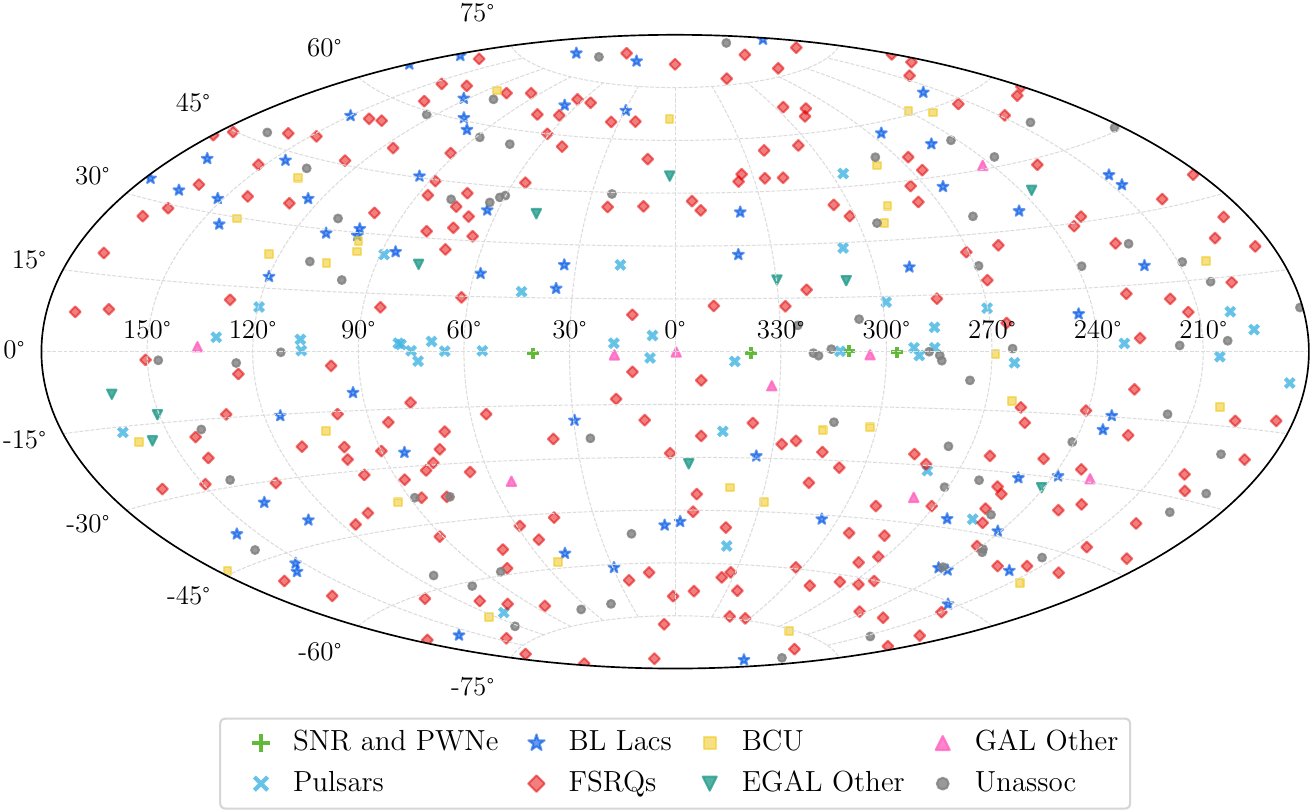}
    \caption{Hammer-Aitoff projection in Galactic coordinates of 2FLE detected sources. Supernova remnants (SNR) and pulsar wind nebulae (PWNe) are indicated with a green cross, pulsars with a cyan cross, BL Lacertae objects (BL Lac, BLL) with a blue star, flat spectrum radio quasars (FSRQ) with a red diamond, blazars of undetermined type (BCU) with a yellow square, while all other extragalactic associations (Other EGAL) are indicated with a green triangle, all other Galactic source types (Other GAL) with a pink triangle, and unassociated sources (Unassoc) with a gray circle. This map showcases the dominance of FSRQs out of the source types detected in the MeV population, as well as preponderance of sources detected at $|b|>10\degree$ in the 2FLE.}
    \label{fig:aitoff_2fle}
\end{figure*}

Finally, we report low-confidence (Bayesian association probability $>$0.1\footnote{Similar to the threshold used in the 4FGL catalog and later releases.}) associations, listed in ASSOC3,  for 45 other sources. For five 2FLE sources located in complex regions of the Galactic plane that suffer from source confusion between point sources and extended ones (e.g. the gamma Cygni region), we also list in ASSOC3 the field of the extended source. 
The breakdown of source associations is outlined in Table \ref{tab:source_assoc_breakdown}.  The best source localization is $0.324\degree$, with 50\% of sources localized better than $0.491\degree$, all at a 95\% confidence level. See Section \ref{sec:effective_psf} for more discussion. 

In general, for the sources' classification, counterpart coordinates and redshift listed in the 2FLE catalog, we follow these criteria: priority is given to the 2FLE associations obtained with the Bayesian or likelihood-ratio (ASSOC1); if ASSOC1 is empty, we use the 4FGL-DR4 counterpart (ASSOC2); if neither ASSOC1 nor ASSOC2 is listed, we rely on the low-probability counterparts (ASSOC3). Coordinates, redshift and classification for the ASSOC1 and ASSOC3 counterparts are taken from the NASA Extragalactic Database (NED\footnote{https://ned.ipac.caltech.edu/}).
For each source associated to a 4FGL source, the best fit 2FLE model with its uncertainty range and its SED is compared to the best fit 4FGL model and its SED as described in Section \ref{sec:sed_constraints}.

\section{2FLE Catalog description} 
\label{sec:2fle_descrip}
\begin{figure*}
    \centering
    \includegraphics[width=1.5\columnwidth]{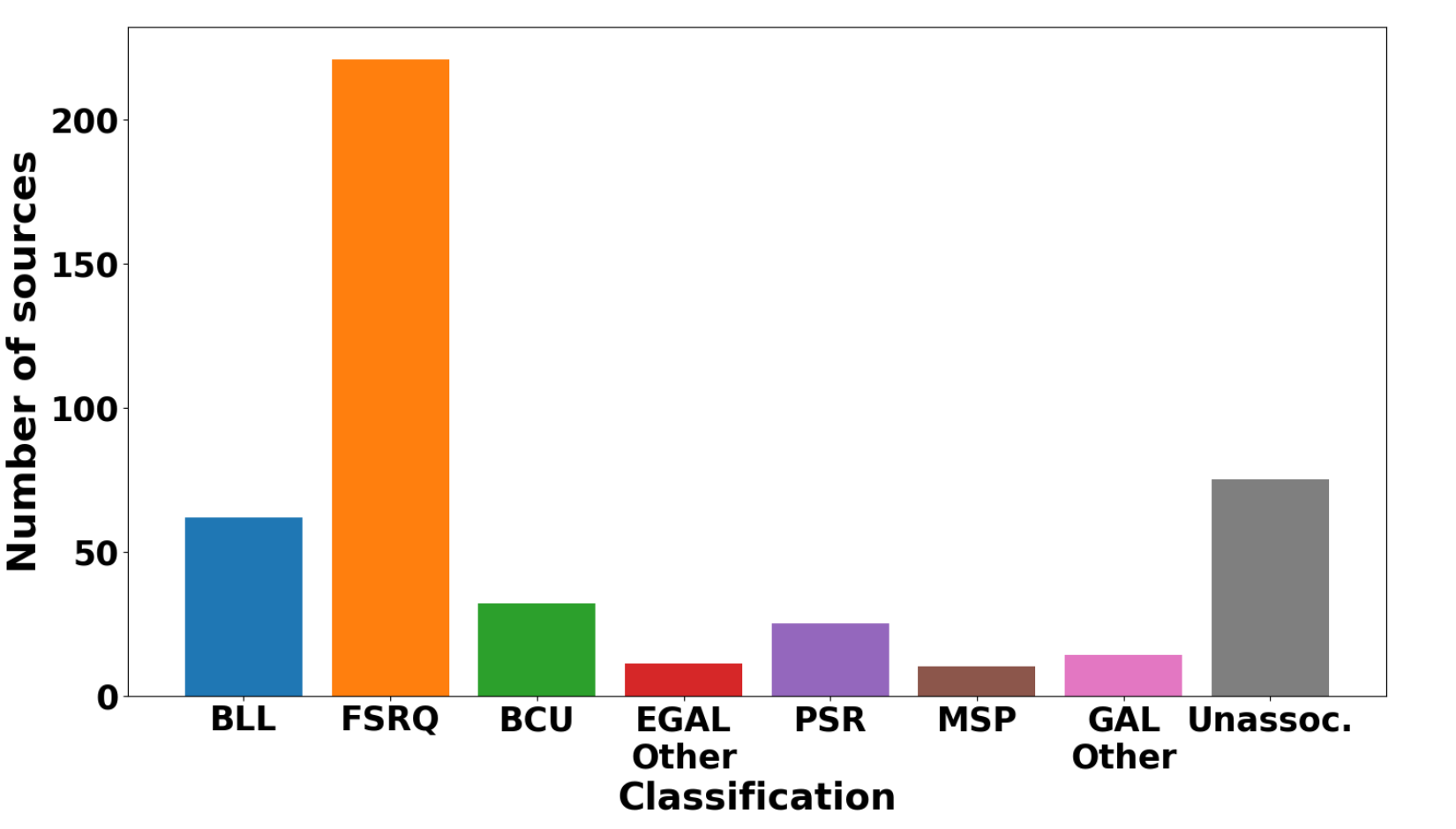}
    \caption{Classifications of 2FLE sources, with unassociated constituting the 73 sources that lack an association. FSRQs dominate the population of extragalactic sources. For Galactic sources, pulsars and millisecond pulsars are the dominant source classes. `EGAL Other' contains the classes of  Narrow-line Seyfert 1 (NLSY1), radio galaxy (RDG), compact steep spectrum quasar (CSS), and active galactic nuclei (AGN). `GAL Other' represents the source classes of a potential association with a supernova remnant or a pulsar wind nebula (SPP), binary (BIN), high-mass binary (HMB), low-mass binary (LMB), and globular cluster (GLC).}
    \label{fig:source_class_distribution}
\end{figure*}

The 2FLE catalog\footnote{\url{https://fermi.gsfc.nasa.gov/ssc/data/access/lat/2FLE/}} contains 450 sources detected with a likelihood $TS>25$. The source designation is 2FLE \texttt{JHHMM.m+DDMM} where 
where HHMM.m is the right ascension in hours and decimal minutes and DDMM is the declination in degrees and minutes. The positions in the catalog are reported in units of degrees for both Galactic and Equatorial (J2000) coordinates. Positional uncertainty regions are reported at the 95\% containment level (see Section \ref{sec:sim_r95}).
Column details of the full catalog are described in Table \ref{tab:cat_col_description}. 
The distribution of 2FLE sources is showcased in an Hammer-Aitoff projection in Figure \ref{fig:aitoff_2fle}. The varying symbols indicate different classes associated with the detected sources based on positional coincidence with corresponding catalogs as described in Section \ref{sec:source_assoc}. 
The breakdown of source classification of the associated counterparts is reported in Table~\ref{tab:assoc_description} and can be visualized in Figure~\ref{fig:source_class_distribution}.
The 2FLE has the majority of sources detected at high Galactic latitudes and 
the dominant source class is that of FSRQs. The source density of the 2FLE at $|b|<20\degree$ is 0.008 sources/$\deg^2$ 
and 0.012 sources/$\deg^2$ 
at $|b|>20\degree$. 
In comparison, the 4FGL source density at $|b|<20\degree$ and $>20\degree$  is 0.25 sources/$\deg^2$ 
and 0.14 sources/$\deg^2$,
 respectively. This is to be expected, as the presence of the diffuse emission, combined with the large LAT PSF at these energies makes detection of objects within the diffuse emission difficult.

To avoid confusion, all association source types use capital letters throughout the paper. To see if the 4FGL reported association is a firm identification please refer to the 4FGL-DR4 catalog. 

\renewcommand{\arraystretch}{1} 
\setlength{\tabcolsep}{1pt}    

\begin{deluxetable*}{cccl}[t]
\centering
   \tablecaption{Catalog Description
    }
    \tablehead{\colhead{Column} & \colhead{Format}  & \colhead{Unit} &\colhead{Description}
    }   
\startdata
\hline
Source\_Name & 18A & & Official Source Name 2FLE JHHMM.m+DDMM\\
SpectrumType & 17A & & Spectral model used (Power Law) \\
RAJ2000 & E & degree & Right Ascension \\
DEJ2000 & E & degree & Declination \\
GLON & E & degree & Galactic Longitude \\
GLAT & E & degree & Galactic Latitude \\
Conf\_95\_SemiMajor & E & degree & Uncertainty radius at 95\% confidence corrected for systematic uncertainties \\
Conf\_95\_SemiMinor & E & degree & SemiMinor axis corrected for systematic uncertainties \\
Conf\_95\_PosAng & E & degree & Positional angle of the fitted ellipse \\
ROI & E & & Region of Interest Number\\
Pivot\_Energy & E & MeV & Energy at which uncertainty on differential flux is minimal \\
Flux\_Density & E & ph cm$^{-2}$ MeV$^{-1}$ s$^{-1}$ & Differential flux at Pivot\_Energy\\
Unc\_Flux\_Density & E & ph cm$^{-2}$ MeV$^{-1}$ s$^{-1}$ & 1$\sigma$ uncertainty on differential flux at Pivot\_Energy \\
Flux & E & ph cm$^{-2}$ s$^{-1}$ & Integral photon flux from $20-200\,\rm MeV$ procured by spectral fitting \\
Unc\_Flux & E & ph cm$^{-2}$ s$^{-1}$ & 1$\sigma$ uncertainty on integral photon flux from $20-200\,\rm MeV$ MeV procured by spectral fitting \\
Flux\_Flag & I & & Set to 1 if the Unc\_Flux $>$ Flux, 0 otherwise\\
Energy\_Flux & E & erg cm$^{-2}$ s$^{-1}$ & Energy flux from 20-200 MeV procured from spectral fitting \\
Unc\_Energy\_Flux & E & erg cm$^{-2}$ s$^{-1}$ & 1$\sigma$ uncertainty on Energy flux from $20-200\,\rm MeV$ procured from spectral fitting \\
PowerLaw\_Index & E & & Best-fit photon number index when fitting with power law \\
Unc\_PowerLaw\_Index & E &  &  1$\sigma$ uncertainty on PowerLaw\_Index \\
Npred & E & & Predicted number of events in the model \\
TS & E & & Test statistic defined by TS = $-$2ln($L_0/L$), comparing the maximum value\\
& & & of the likelihood function including the source in the model $L$ to the value without it $L_0$\\
Flux\_Density\_Band & E & ph cm$^{-2}$ MeV$^{-1}$ s$^{-1}$ & Differential photon flux evaluated in each spectral band at the bin central energy ($E_{\rm mid}$, Sec.~\ref{sec:sed_gen}) \\
Unc\_Flux\_Density\_Band & E & ph cm$^{-2}$ MeV$^{-1}$ s$^{-1}$ & $1\sigma$ lower and upper error on Flux\_Density\_Band\\
TS\_Band & E & & Test Statistic in each spectral band\\
ASSOC\_GAM & 18A & &  4FGL-DR4 counterpart\\
TEVCAT\_FLAG & A & & P if positional association with non-extended source in TeVCAT,\\
& & & E if associated with an extended source in the TeVCAT\\
& & & N if no TeV association\\ 
ASSOC\_TEV & 26A & & Name of likely corresponding TeV source from TeVCAT, if any\\
CLASS & 5A & & Class designation for associated sources; see Table \ref{tab:assoc_description}\\
ASSOC1 & 26A & & Name of identified or likely associated source \\
ASSOC2 & 26A & & Name of source associated with the ASSOC\_GAM source \\
ASSOC3 & 26A & & Name of low-confidence counterpart \\
ASSOC\_PROB & E & & The probability of association based on the Bayesian method and on the Likelihood Ratio\\ &&& for 2FLE J0340.1+3215 and 2FLE J1458.1-3602\\
ASSOC\_PROB\_BAY\_DR4  & E & &  The probability of association  based on the Bayesian method  for the ASSOC\_GAM source\\ 
& & &  except when ASSOC2 is not filled   \\
RA\_Counterpart & D & degree & 
The Right Ascension of the counterpart reported in the ASSOC1 columns\\
& & & or ASSOC2 if ASSOC1 is not filled, ASSOC3 otherwise \\
DEC\_Counterpart & D & degree & 
The Declination of the counterpart reported in the ASSOC1 columns\\
& & & or ASSOC2 if ASSOC1 is not filled, ASSOC3 otherwise \\
Redshift & E & & Redshift of counterpart if known \\
Redshift\_Flag & 4A & & Quality flag for known heliocentric redshift\\
& & & See \url{https://ned.ipac.caltech.edu/Documents/Guides/Database} for flag meaning\\
\enddata
\end{deluxetable*}\label{tab:cat_col_description}

\subsection{General Characteristics of 2FLE Sources}

\begin{figure}
    \centering
    \includegraphics[width=1\columnwidth]{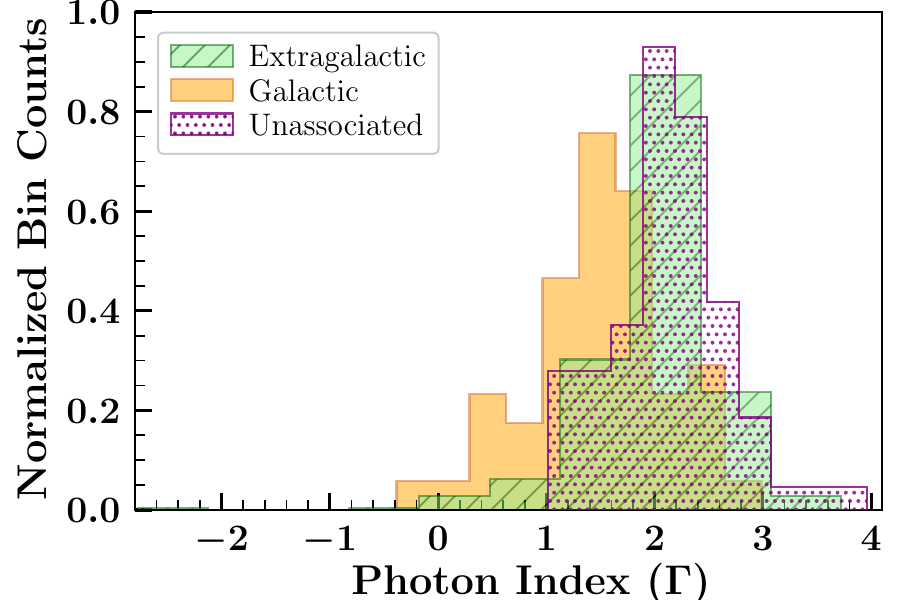}
    \caption{Photon Index distribution of Galactic (orange, solid), extragalactic (green, diagonal lines), and unassociated (purple, dotted) sources. In general, the unassociated source index primarily overlaps with extragalactic sources' photon index. }
    \label{fig:2fle_index_comparison_egal_gal_unassoc}
\end{figure}

\begin{figure}
    \centering
    \includegraphics[width=1\columnwidth]{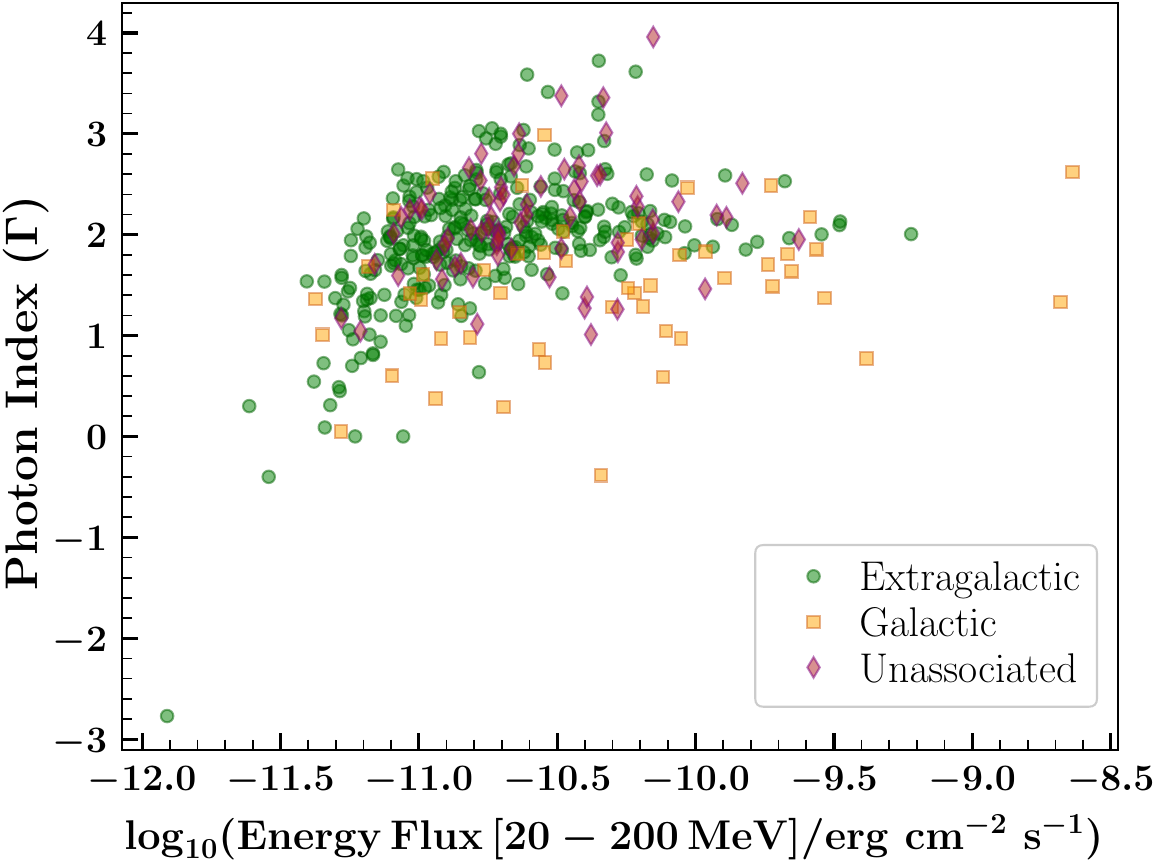}
   \caption{Photon index versus energy flux between 20 and 200 MeV of 2FLE detected sources. Extragalactic (green circles), Galactic (orange squares) and unassociated (purple diamonds) sources are plotted separately. 
   The absence of sources in the top left corner reflects the dependence
   of the sensitivity on photon index.   }
    \label{fig:2fle_flux_vs_gam}
\end{figure}

The 2FLE contains 377 sources associated with a multiwavelength counterpart (Section~\ref{sec:source_assoc}) and 73 unassociated sources (UNASSOC). Out of these 377 sources, 325 are associated with sources of extragalactic (EGAL) origin and 52 with objects of Galactic (GAL) origin.
All sources were modeled as a power law, and the resulting median photon index is $\tilde{\Gamma} \sim 2$.
Figure \ref{fig:2fle_index_comparison_egal_gal_unassoc} showcases that Galactic associated sources tend to have harder photon indices than extragalactic associated and unassociated sources ($\tilde{\Gamma}_{\rm GAL} = 1.48$, $\tilde{\Gamma}_{\rm EGAL} = 1.99$, $\tilde{\Gamma}_{\rm UNASSOC}=2.17$). The contrast between Galactic and extragalactic photon indices likely points to intrinsically different emission mechanisms in Galactic objects, such as synchrotron emission and pion decay. Moreover, the unassociated sources have on average the softest photon index in the 2FLE, which could explain why they remain undetected in the higher energy band covered by the 4FGL.
Figure \ref{fig:2fle_flux_vs_gam} compares the 2FLE photon index against the energy flux for each of the association categories (extragalactic, Galactic, and unassociated). The figure indicates that 
the minimal detectable energy flux depends on the photon index;
because of the energy-dependent effective area, harder sources can be detected to lower energy fluxes than soft-spectrum sources. 

\begin{figure}
    \centering
    \includegraphics[width=1\columnwidth]{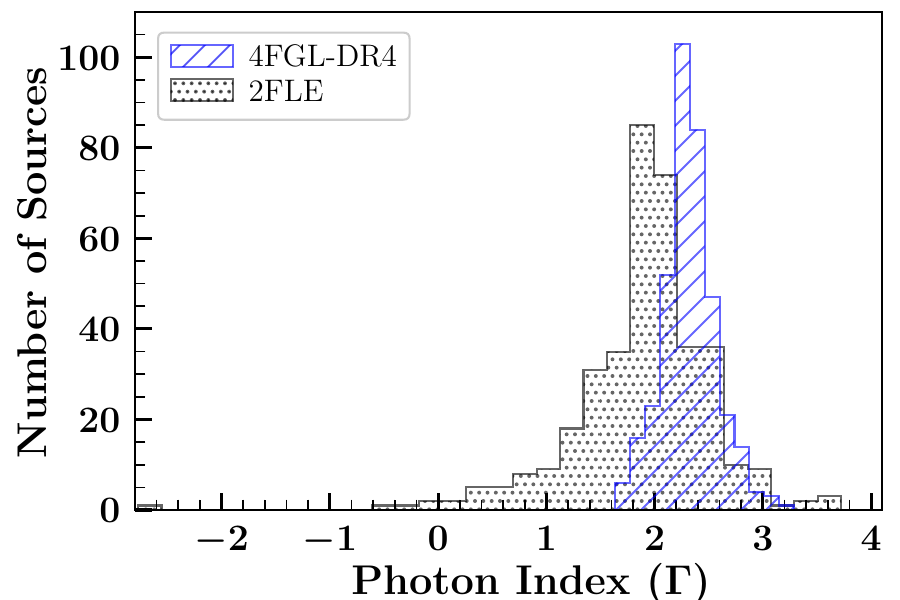}
    \caption{Photon index distribution of 2FLE sources (black, dotted) associated with a 4FGL one and their corresponding 4FGL-DR4 photon indices (blue, hatched). The 20-200 MeV energy band is critical for sampling the peak emission of these sources, as seen by the 2FLE index distribution peaking around $\Gamma=2.0$ compared to the softer indices reported by their 4FGL associations. }
    \label{fig:2fle_index_vs_4fgl_NOTLOCAL}
\end{figure}

\begin{figure*}
    \centering
    \includegraphics[width=0.9\textwidth]{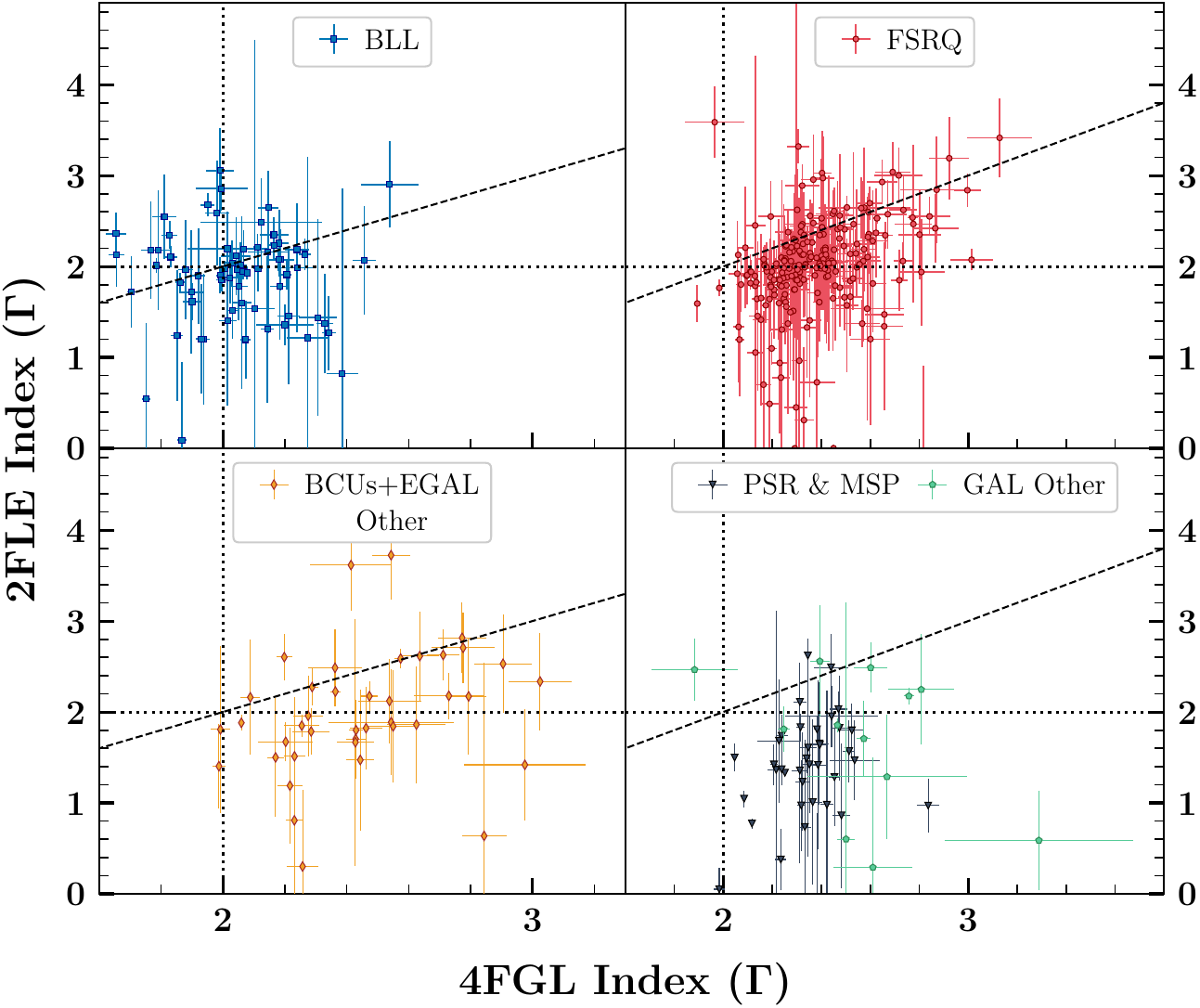}
    \caption{2FLE and 4FGL best fit spectral index value for the 2FLE sources associated with: BLL (light blue squares, top left), FSRQ (red points, top right), BCUs \& EGAL Other (yellow diamonds, bottom left), PSR \& MSP (dark blue triangles), and GAL Other (light green pentagons) on the bottom right. EGAL Other and GAL Other contain the same source classes as reported in Figure \ref{fig:source_class_distribution}. The dashed black line is the one-to-one line. The dotted black lines indicate the transition from a hard to soft spectrum ($\Gamma=2$).}
    \label{fig:2fle_index_vs_4fgl_src_compare}
\end{figure*}

Figure \ref{fig:2fle_index_vs_4fgl_NOTLOCAL} showcases the photon index reported by 2FLE and the 4FGL for associations between the two catalogs. 
The 2FLE generally reports harder photon indices than the 4FGL ($\tilde{\Gamma}_{\rm 2FLE}=1.96$ vs.~$\tilde{\Gamma}_{\rm 4FGL}=2.31$), indicating that in the MeV band the 2FLE samples the SEDs of the sources either closer to the peak of the emission, or in the rising part of the SED. 
Figure \ref{fig:2fle_index_vs_4fgl_src_compare} plots the one-to-one correspondence of the 2FLE and 4FGL reported spectral index by source class for the same associated 2FLE objects reported in Figure \ref{fig:2fle_index_vs_4fgl_NOTLOCAL}.

The median energy flux of the 2FLE sources is, $\tilde{F}_{20-200 \, \text{MeV}} = 1.90\times10^{-11}$ erg cm$^{-2}$ s$^{-1}$, with the minimum and the maximum being $F_{20-200 \,\rm MeV,~min} = 1.23\times10^{-12}$ erg cm$^{-2}$ s$^{-1}$, and $F_{20-200 \, \rm MeV,~max} = 2.30\times10^{-9}$ erg cm$^{-2}$ s$^{-1}$ , respectively. Furthermore, of the top 25 brightest sources by energy flux in the entire catalog, 11 are Galactic, 10 are extragalactic, and 4 are unassociated.

\renewcommand{\arraystretch}{1} 
\setlength{\tabcolsep}{1pt}         

\begin{deluxetable*}{lccccc}[t]
\centering
   \tablecaption{Classification of 2FLE sources with a multi-wavelength association} 
    \tablehead{\colhead{4FGL Description} & \colhead{} & \colhead{Designator} & \colhead{Number} & \colhead{Designator}& \colhead{Number}
    }   
\startdata
\hline
Non-blazar active galaxy &  & AGN & 1 & agn & 0 \\
BL Lac type of blazar    &  & BLL  & 17 & bll  & 50 \\
FSRQ type of blazar      &  & FSRQ & 28 & fsrq & 190 \\
Blazar candidate of uncertain type  &  & BCU & 0 & bcu  & 29 \\
Narrow-line Seyfert 1  &  & NLSY1 & 3 & nlsy1 & 2 \\
Radio galaxy           &  & RDG & 2 & rdg  & 2 \\
Compact steep spectrum quasar & & CSS & 0 & css & 1 \\
High-mass binary  &  & HMB & 3 & hmb & 0 \\
Low-mass binary   &  & LMB & 2 & lmb & 0 \\
Binary            &  & BIN & 0 & bin & 1 \\
Pulsar, previously identified by pulsations  &  & PSR &  29 & psr & 0 \\
Millisecond Pulsar  &  & MSP & 10 & msp & 0 \\
Globular cluster    &  & GLC & 0 & glc & 2 \\
Supernova remnant (SNR) / Pulsar wind nebula (PWN)  &  & SPP/PWN &  0 & spp &  4 \\
Galactic Center  &  & GC & 1 & -- & --\\
\hline
\enddata
\tablecomments{We follow the 4FGL nomenclature for classifications, where upper case letters denote a firm identification, while lowercase letters denote an association. The number reported to the right of each `Designator' column is indicating the number of sources with that specific designation in the 2FLE. The criteria for reporting source classification are described in Section~\ref{sec:source_assoc}. }
\end{deluxetable*}\label{tab:assoc_description}

\begin{figure}
    \centering
    \includegraphics[width=1\columnwidth]{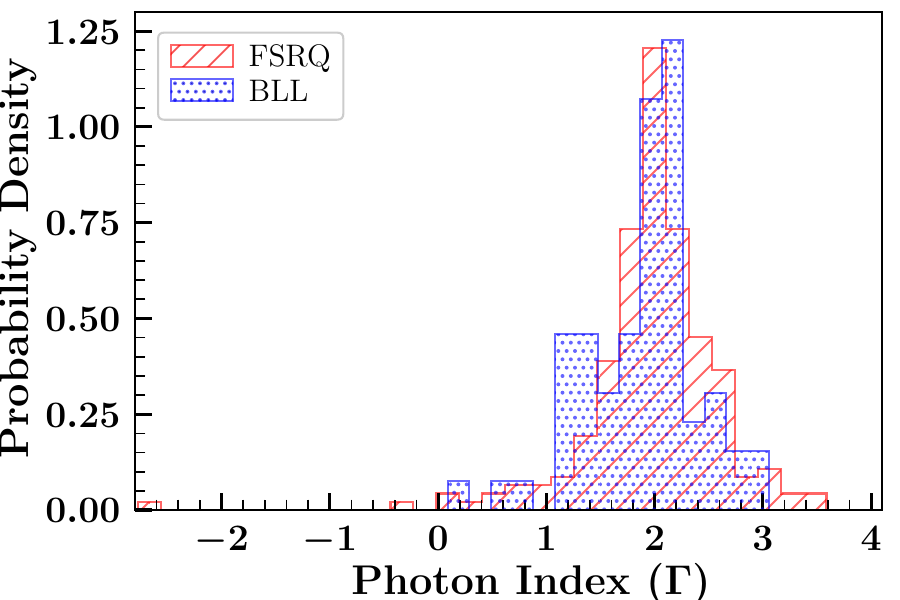}
    \caption{ Distribution of photon indices for the populations of FSRQs (red) and BLLs (blue) as reported in the 2FLE. Not only did the 2FLE detect MeV-dominated FSRQs, but also low-, intermediate- and high-synchrotron peaked BLLs, the latter displaying hard MeV photon indices hence likely peaking in the GeV--TeV band.
    We note that the distribution of BLLs in index is asymmetric, with more BLLs reported with indices of $\Gamma<2$ compared to $\Gamma>2$.}
    \label{fig:fsrq_vs_bll_index}
\end{figure}

\begin{figure}
    \centering
    \includegraphics[width=1\columnwidth]{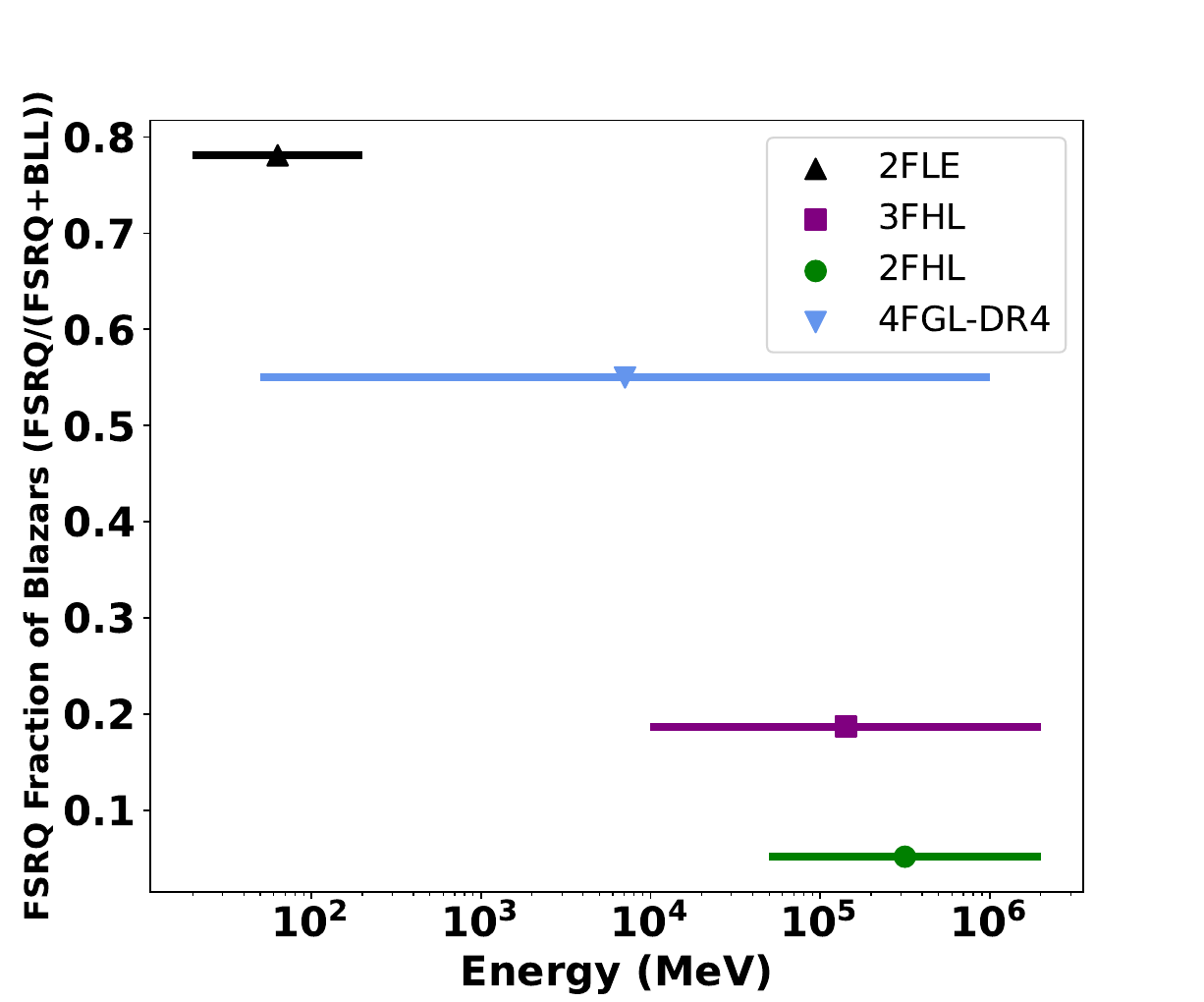}
    \caption{Fraction of FSRQs (FSRQ / (FSRQ + BLL)) plotted against the energy range for each catalog. Clearly, the softer gamma-ray catalogs detect a higher fraction of FSRQs, with the 2FLE reporting the highest fraction out of the 2FLE, 4FGL-DR4, 2FHL, and the 3FHL. This indicates that most FSRQs peak emission occurs in the MeV, while generally BLLs peak at GeV and TeV energies}
    \label{fig:fsrq_fraction}
\end{figure}

\begin{figure}
    \centering
    \includegraphics[width=1.1\columnwidth]{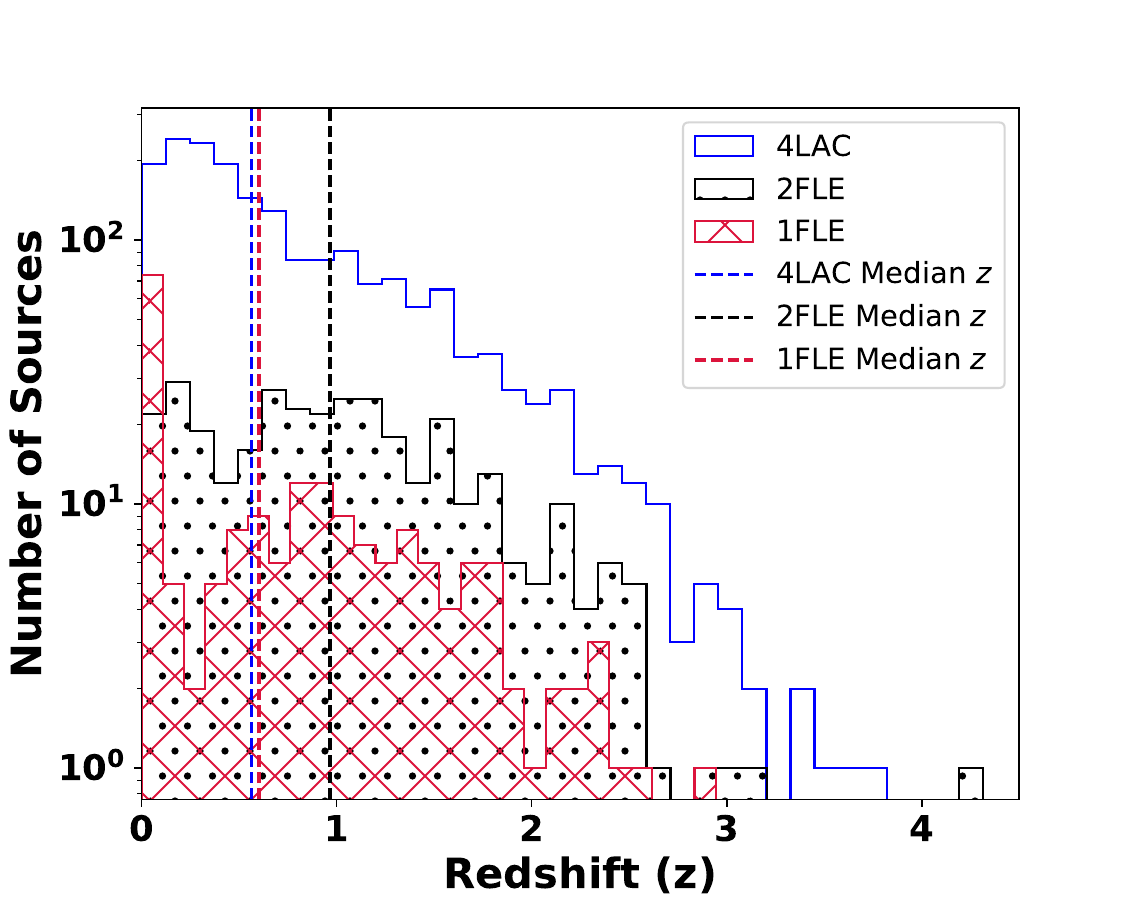}
    \caption{Distribution of source redshift in the 4LAC (blue), 2FLE (black, dotted) and 1FLE (red, cross-hatched) with corresponding median values indicated by dashed vertical lines. The 2FLE detects the largest median redshift value and has a flatter distribution of sources detected at $0.6<z<1.5$ compared to the 1FLE and 4LAC. The dominance of FSRQs in the 2FLE is the primary reason for this difference. The preponderance of FSRQs in the 2FLE is generally due to FSRQs high-energy spectral hump peaking at MeV energies, especially FSRQs at higher $z$. We note that the $z=4.314$ source is also detected in the 4LAC.}
    \label{fig:redshift}
\end{figure}

\begin{figure*}
    \centering
    \includegraphics[width=\textwidth]{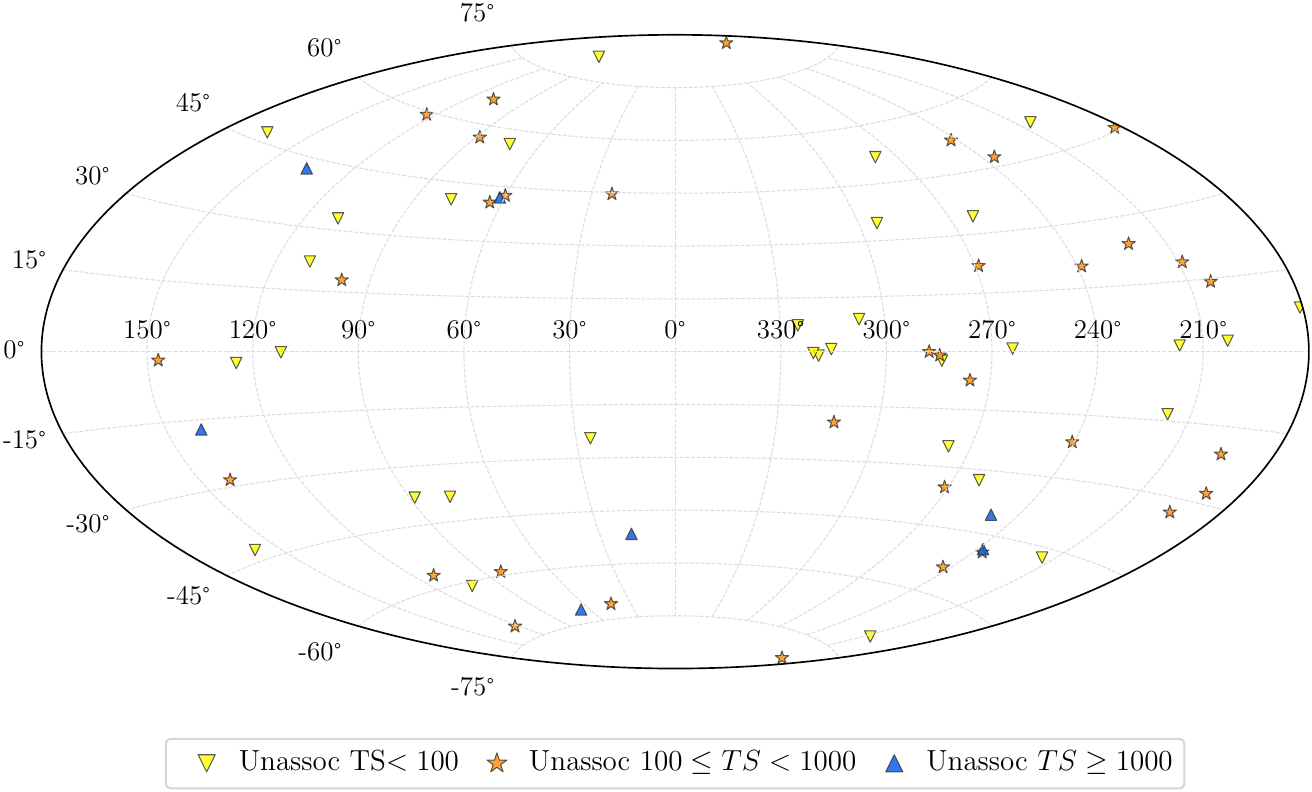}
    \caption{Hammer-Aitoff projection in Galactic coordinates of unassociated 2FLE detected sources. Downward pointing yellow triangles correspond to unassociated sources detected at $TS<100$, red-orange stars at $100<TS<1000$, and upward pointing blue triangles to unassociated sources detected at $TS>1000$. The majority of $TS>1000$ sources are found at high Galactic latitudes. We note a concentration of unassociated, $TS<100$ sources detected at low-latitudes between a Galactic longitude of $300-330\degree$.  }
    \label{fig:aitoff_unassoc_2fle}
\end{figure*}

\subsection{Extragalactic Sources}\label{sec:egal}

The 2FLE includes 325 sources associated with extragalactic sources. The extragalactic source with the highest $TS$ value is 2FLE J2253.7+1608 ($TS=15794$). This source is associated with 3C 454.3, the brightest known $\gamma$-ray blazar in all 4FGL catalogs \citep{4lac-dr3, 4fgl-dr4}. 3C 454.3 is also detected with the highest energy flux of the extragalactic sources ($6.0 \pm 0.1\times 10^{-10}$ erg cm$^{-2}$ s$^{-1}$). The source classifications detected are: flat spectrum radio quasars (FSRQ, 218 sources), BL Lacertae objects (BLLs, 67 sources), blazars of uncertain type (BCU, 29), Narrow Line Seyfert 1 Galaxies  (NLSY1, 5 sources), radio galaxies (RDG, 4 sources), compact steep spectrum quasar (CSS, 1 source), and other active galactic nuclei (AGN, 1 source associated with PKS 0521-36, \citealp[e.g.,][]{dammando_2015}). FSRQs are the dominant source class, with 218/325 ($\sim67\%$) of the extragalactic associated sources and 49\% of the entire catalog. See Table \ref{tab:assoc_description} for a full description of source associations.
Of the extragalactic associated sources, 95\% are found at a Galactic latitude of $|b|\geq 10\degree$.
Only 14 are detected at low Galactic latitudes ($|b|<10\degree$), of which 11 are FSRQs.

The low-energy hump of blazars' broadband SED (peaking anywhere between mid-IR to X-rays) is well explained by synchrotron emission of electrons. One way to classify blazars is based on the peak synchrotron emission frequency ($\nu^s_{peak}$). These classifications are  high ($\nu^s_{peak}\geq10^{15}$\,Hz), intermediate ($ 10^{14} \leq \nu^s_{peak}\leq10^{15}$\, Hz), and low ($\nu^s_{peak}\leq10^{14}\,$Hz) synchrotron peaked blazars (HSP, ISP, LSP; \citealp{abdo2010spectral}). The positions of the synchrotron peak of the 2FLE's 4FGL associations are reported in the Fourth Catalog of Active Galactic Nuclei detected by the \textit{Fermi} Large Area Telescope: Data Release 3 (4LAC-DR3, \citealp{4lac-dr3}) for 66 out of the 67 2FLE BLLs. There are 39 LSPs, 13 ISPs and 14 HSPs for the BLL population. Thus, HSPs are a minority. No detected BLL qualifies as extreme. The most extreme cases present in the 2FLE are Markarian 501 (2FLE J1655.9+3942) and Markarian 421 (2FLE J1105.2+3809).
On the other hand, among the associated 2FLE FSRQ population, we find that practically all are LSPs. This is expected since if most FSRQs have an inverse-Compton emission peak in the MeV, their synchrotron peak must be at lower frequencies. 
The high-energy hump of the blazar SED (peaking anywhere between MeV to TeV energies) is predicted to originate either from leptonic inverse Compton scattering
\citep{sikora1994comptonization_EC}, or alternatively from lepto-hadronic processes \citep{Bottcher_2013_lep_had_model,Murase_2016_hidden,costamante2019blazars}. This is the portion of the blazar SED that the 2FLE captures.

Looking at the distribution of indices for the 2FLE population of FSRQs and BLLs (as shown in Figure \ref{fig:fsrq_vs_bll_index}), we note that both source classes peak around $\Gamma\sim2$ ($\tilde{\Gamma}_{\rm 2FLE,\,BLL}=1.97, \tilde{\Gamma}_{\rm 2FLE,\,FSRQ}=2.02$). On the other hand, the 4FGL-DR4 reports a median photon index of $\tilde{\Gamma}_{\rm4FGL,\, BLL}=2.0$, $\tilde{\Gamma}_{\rm 4FGL,\, FSRQ}=2.47$ for BLLs and FSRQs, respectively. As depicted in Figure \ref{fig:2fle_index_vs_4fgl_src_compare}, the 2FLE is detecting the peak of high-energy emission of FSRQs sources as opposed to the falling edge in the 4FGL.

In comparison to the previous $\gamma$-ray catalogs -- i.e.~the Second Catalog of Hard \textit{Fermi}-LAT Sources \citep[2FHL,][]{2FHL}, the third Catalog of Hard \textit{Fermi}-LAT Sources \citep[3FHL,][]{3FHL}, and the 4FGL-DR4 -- we detected the highest percentage of FSRQs as a fraction of the cumulative known blazar types (FSRQ + BLL) at 78\%. Figure \ref{fig:fsrq_fraction} depicts this fraction against the energy range of the 2FLE, 4FGL-DR4, 2FHL, and 3FHL. This reinforces the dominance of FSRQ emission at MeV energies compared to BLLs being more dominant at GeV and TeV energies. The positions of the 4FGL inverse-Compton peaks (based on the \texttt{LogParabola} fits to the 4FGL SEDs) are given in Fig. 6 of the 4LAC-DR3 paper (\citealp{4fgl_dr3}). For FSRQs, they are clustered below 1 GeV while BL Lacs generally peak at larger energies.

Based on the source associations discussed in Section \ref{sec:source_assoc}, in Figure \ref{fig:redshift} we report the distribution of sources with a known redshift compared to the 1FLE and the 4LAC-DR3. Of the three catalogs, the 2FLE reports the highest median redshift value, of $\tilde{z}_{2FLE} = 0.966$. The median redshift value reported by the 4LAC is $\tilde{z}_{4LAC} = 0.563$, and the 1FLE is $\tilde{z}_{1FLE} = 0.602$. 
The 2FLE detects more sources at higher redshifts than the 1FLE, but the 4LAC detects a larger number of high-$z$ sources. This is likely due to the sensitivity limitations of \textit{Fermi}-LAT at MeV energies. The 2FLE reports 339 sources with redshifts. The majority of associated 2FLE sources are at $z<2$, with 87GB 150844.6+571424 ($z=4.314$) being the highest redshift source detected.

\subsection{Galactic Sources}\label{sec:GAL}
The 2FLE comprises 52 sources associated with Galactic objects. The source populations represented in the 2FLE are high-mass binaries (HMB, 3 sources), low-mass binaries (LMB, 2 sources), binaries (BIN, 1 source), globular clusters (GLC, 2 sources), millisecond pulsars (MSP, 10 sources), young pulsars (PSR, 29 sources), potential supernova remnants/pulsar wind nebulae (SPP/PWN, 4 sources), and one source associated with the Galactic Center (GC). Table \ref{tab:assoc_description} contains which of these sources are reported as identified classifications in the 4FGL and which are simply reported as an association. 

Of the Galactic sources, 34 (65\%) are found at Galactic latitudes of $|b|<10\degree$. Half of the high-latitude Galactic sources are millisecond pulsars. The globular cluster M2 which was not detected in any previous \textit{Fermi}-LAT catalog is also at a higher galactic latitude. The most significantly detected source is the Crab pulsar (2FLE J0534.3+2159) at TS$=23013$ which is also the brightest source in the catalog with an energy flux of $2.30\pm 0.03\times10^{-9}$ erg cm$^{-2}$ s$^{-1}$.

\subsection{Unassociated Sources}\label{sec:unassoc}
The catalog reports 73 unassociated sources ($\sim16\%$ of the catalog). Of these, 60 are newly discovered gamma-ray sources, while 13 are detections of previously unassociated 4FGL sources. Figure \ref{fig:aitoff_unassoc_2fle} is a Hammer-aitoff projection on the sky with only the unassociated sources displayed, differentiating sources detected at $TS<100$ (32), $100<TS<1000$ (34) and $TS>1000$ (7, see Table~\ref{tab:2flets1000}).   
We anticipate that most of the unassociated sources will be of extragalactic origin. With 87\% of the associated sources being of extragalactic origin, a naive assumption would be that $\sim62$ of the unassociated sources would also be extragalactic. This number is likely higher due to the unassociated source median photon index being similar to that of the extragalactic population ($\tilde{\Gamma}_{\rm UNASSOC}=2.17$, Figure~\ref{fig:2fle_index_comparison_egal_gal_unassoc}).  

As for their distribution in Galactic latitude, among the 73 unassociated sources, 36 are located at 
high Galactic latitude ($|b|>35\degree$). Thus, almost half of the unassociated population is found at high latitudes.
Moreover, 16 of those (22\%) are found at low latitudes ($|b|<10\degree$), including 12/16 (75\%) at very low latitudes ($|b|<5\degree$).
Since these low latitude sources have a wide range of photon indices, these cannot be associated with a specific population of sources. As for their distribution in energy flux, these very low latitude sources have a median 20-200 MeV energy flux of $5.8\times 10^{-11}$ erg cm$^{-2}$ s$^{-1}$, which is greater than 81\% of the unassociated sources in the catalog. This is a likely selection effect as sources with low fluxes would not make the TS cut in these regions with high diffuse emission.
Furthermore, there are 4 unassociated sources at $|b|<5\degree$ which also have $\Gamma<1.49$, which could potentially indicate a Galactic nature. Multiwavelength campaigns are necessary to determine the source classes, and origin, of these unassociated sources.

We also note that it is possible that some of the unassociated sources are unassociated due to source confusion and blending effects from the large PSF at these energies. One such example is 2FLE J1050.8+7125. Found to be unassociated using our procedure in Section \ref{sec:source_assoc}, but is likely related to 4FGL J1048.4+7143 (associated with the FSRQ S5 1044+71) whose position is offset due to the proximity of other nearby fainter sources. Another case is 2FLE J0220.3+4248, whose position falls in between two MeV bright 4FGL sources (4FGL J0222.6+4302 associated with the BL Lac 3C 66A; and 4FGL J0218.1+4232 associated with the millisecond pulsar PSR J0218+4232). In Table~\ref{tab:2flets1000} we list all the $TS>1000$ 2FLE sources, which are likely suffering from similar blending issues rather than being newly detected MeV sources. Robust follow-up investigating source confusion is left for a future work. 

\begin{deluxetable*}{lccccc}[t]
\centering
   \tablecaption{List of $TS>1000$ 2FLE unassociated sources.}
    \tablehead{\colhead{Source Name} & \colhead{ TS$_{\rm 2FLE}$ } & \colhead{4FGL neighbor\tablenotemark{a}} & \colhead{TS$_{\rm 4FGL, 50-300\,\rm MeV}$\tablenotemark{b}} & \colhead{ASSOC$_{\rm 4FGL}$} & \colhead{ TYPE$_{\rm 4FGL}$} \\
    }   
\startdata
\toprule
    2FLE J1636.8+3833 & 3257 & 4FGL J1635.2+3808 & 1001 & 4C +38.41 & FSRQ \\
    \midrule
    2FLE J1050.8+7125 & 1809 & 4FGL J1048.4+7143 & 668 & S5 1044+71 & FSRQ  \\    
    \midrule
    2FLE J0446.8$-$4357 & 1331 & 4FGL J0449.4-4350 & 104 & PKS 0447-439 & BL Lac \\   
                        &      &  4FGL J0440.3-4332 & 157 & PKS 0438-43 &  FSRQ \\   
    \midrule
    2FLE J2155.7$-$3021 & 1294 & 4FGL J2158.8-3013 & 651 & PKS 2155-304 & BL Lac \\  
    \midrule
    2FLE J2346.3-1612 & 1291 & 4FGL J2345.2-1555 & 281 &  PMN J2345-1555 & FSRQ \\
                        & & 4FGL J2348.0-1630 & 233 &  PKS 2345$-$16 & FSRQ\\
    \midrule
      2FLE J0405.0$-$3623 & 1182 & 4FGL J0403.9-3605\tablenotemark{c} & 3184 & PKS 0402-362 & FSRQ \\
    \midrule
      2FLE J0220.3+4248 & 1008  & 4FGL J0222.6+4302 & 308 & 3C 66A & BL Lac \\
                        & & 4FGL J0218.1+4232 &  240 & PSR J0218+4232 & MSP \\
    \bottomrule
\enddata
    \tablenotetext{a}{4FGL MeV bright sources ($TS_{\rm 4FGL, 50-300\,\rm MeV}>100$) just outside the 2FLE $R_{95}$ (Equation~\ref{eq:r95}).}   
    \tablenotetext{b}{Sum of the squares of the source significance in the first two SED bands listed in the 4FGL catalog \citep{4fgl-dr4}.}  
    \tablenotetext{c}{The MeV emission from 4FGL J0403.9-3605 may have been ``split" in two between 2FLE J0405.0$-$3623 and 2FLE J0401.5$-$3535 (TS=104) on the other side of the 4FGL position.}
\end{deluxetable*}\label{tab:2flets1000}

\begin{figure}[h!]
\includegraphics[width=0.95\columnwidth]{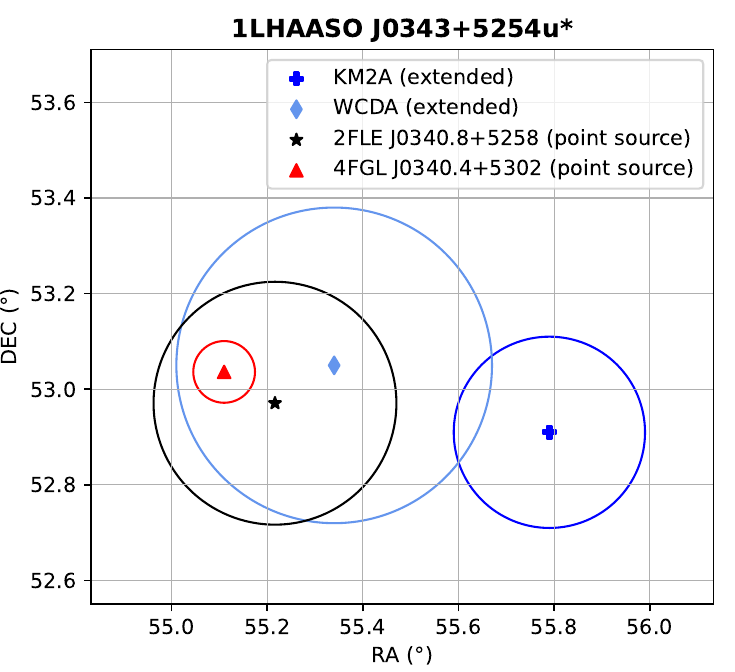}
    \caption{Positional coincidence of LHAASO (KM2A, dark blue plus; WCDA light blue diamond), 2FLE (black star), and 4FGL (red triangle) of the unassociated source 2FLE J0340.8+5258.}
    \label{fig:2fle_j0343_tev}
\end{figure}

\subsection{TeV Coincident Sources}
As discussed in Section \ref{sec:intro}, the MeV band sampled by the 2FLE could potentially help identify SNR which contain a hadronic ``pion bump'' as well as synchrotron emission produced by MeV pulsars and their PWN environments. A shared trait that is possible for both of these source types and their surroundings is the presence of extended TeV emission. Therefore, searching for potential counterparts in the TeV and GeV regime may help to identify new supernova remnants/pulsar wind nebulae.

Based on the association procedure in Section~\ref{sec:source_assoc}, using the TeV catalogs \citep{horan2004extragalactic, chang20172whsp},
45 2FLE sources have a TeV association, 10 with an extended TeV source, and 35 with a point source (out of the searched catalogs). 
One catalog dedicated to the investigation of identifying extended TeV emission with a blind search is the First LHAASO Catalog of Gamma-Ray Sources (1LHAASO, \citealp{LHAASO}), a catalog that was not considered in the association procedure. LHAASO contains two detectors, the Water Cherenkov Detector Array (WCDA, $E=1$\,TeV$-30$\, TeV) and Kilometer Squared Array (KM2A, $E>25$\, TeV). After cross-matching the positions of the 2FLE and 1LHAASO sources, we find 18 2FLE sources are positionally coincident with a LHAASO source's 39\% containment radius of the 2D-Gaussian model ($r_{39}$) used to model the LHAASO source's extension. All of these 18 sources are positionally coincident with a 4FGL source, 6 of which do not have a corresponding TeV counterpart in the main 2FLE catalog. 
Only one, 2FLE J0340.8+5258 (4FGL J0340.4+5302), lacks an associated optical/infrared counterpart as well (see Section~\ref{sec:source_assoc}). 
More details on this source are discussed below and Table \ref{tab:lhaaso_2fle_comparison} outlines the characteristics of potential TeV counterparts of the 6 2FLE sources lacking a TeV association. 

\subsubsection{2FLE J0340.8+5258}
\begin{figure*}
    \centering
    \includegraphics[width=\textwidth]{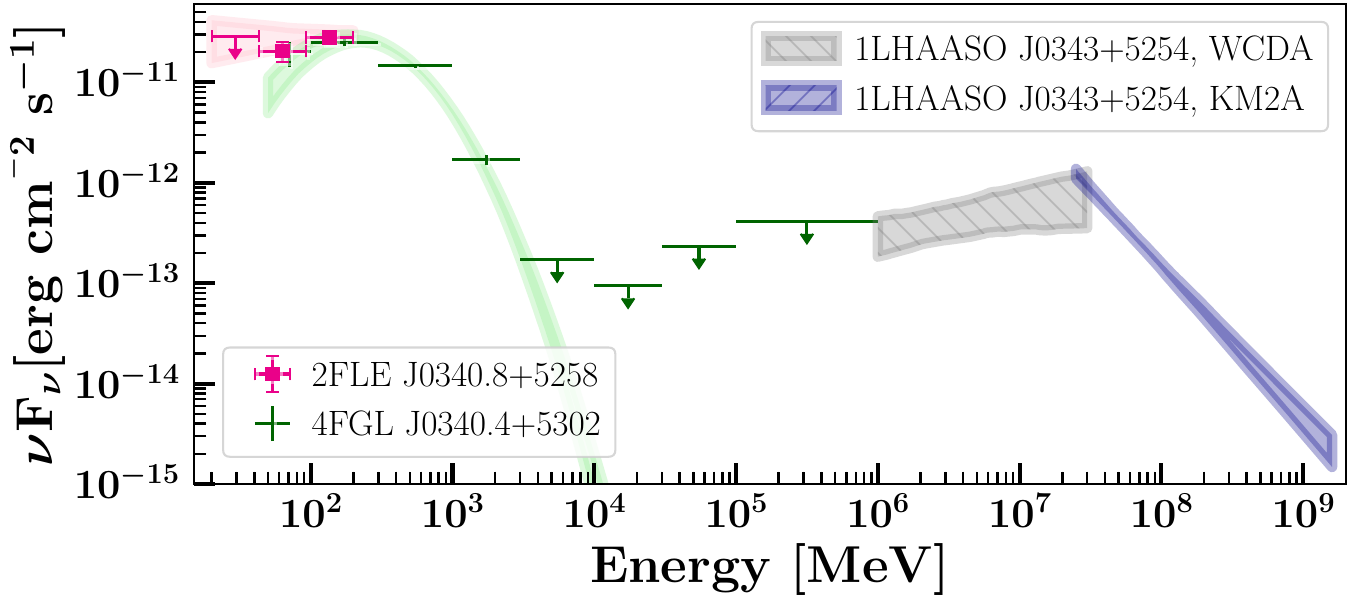}
    \caption{20 MeV to 1.6 PeV SED of 2FLE J0340.8+5258. The 2FLE SED data points (magenta squares) are computed according to Section~\ref{sec:sed_gen} and the 4FGL data is extracted from the 4FGL-DR4 catalog. The shaded bands represent the $1\sigma$ confidence levels obtained using the best-fit parameters of the source from the 2FLE (magenta), 4FGL-DR4 (green) and LHAASO WCDA (gray, backward hatch) and KM2A (blue, forward hatch) parameters as listed in \citet{LHAASO}.}
    \label{fig:lhasso_sed}
\end{figure*}

2FLE J0340.8+5258 is positionally coincident with 4FGL J0340.4+5302, an unassociated source found at a low Galactic latitude ($b=-1.84\degree$). 2FLE J0340.8+5258 is found to be positionally coincident with the WCDA and KM2A detection of 1LHAASO J0343+5254u* \footnote{$u$ indicates that the source is an ultra-high-energy (UHE) source with $TS>20$ for $E>100$ TeV and * denotes a detected source that is suspected of originating from two sources (\citealp{LHAASO})} (see Figure \ref{fig:2fle_j0343_tev}). In Figure~\ref{fig:lhasso_sed} we show the source SED, from 20 MeV to 1.6 PeV. In the 4FGL, the source spectrum is well described by a log-parabola spectrum with photon index $\Gamma_{\rm 4FGL}=3.34\pm0.06$ and peaks at $E_{\rm peak}=202\pm18\,\rm MeV$. The 2FLE harder power-law photon index $\Gamma_{\rm 2FLE}=1.95\pm0.22$ agrees with the idea that in the MeV we are sampling the source's spectral energy distribution close to the peak of the emission, rolling over in the GeV band. Analogously, in the very high-energy regime sampled by LHAASO, we observe the same rise and fall of the spectral energy distribution with similar photon indices agreeing in the low-energy counterpart, with $\Gamma_{\rm WCDA}=1.7\pm0.9$ and $\Gamma_{\rm KM2A}=3.5\pm0.1$, possibly suggesting a connection between these two peaks, and strengthening the association between the MeV and the TeV source. 

\renewcommand{\arraystretch}{0.9} 
\setlength{\tabcolsep}{2pt}         

\begin{deluxetable*}{lcccc|lcclcl}[t]
\centering
\tabletypesize{\footnotesize} 
\tablecolumns{11} 
\tablecaption{2FLE TeV Unassociated Sources Spatially Coincident with a LHAASO Detection}
\tablehead{ \multicolumn{5}{c|}{LHAASO\tablenotemark{a}} & \multicolumn{6}{c}{2FLE\tablenotemark{b}}  \\
\colhead{Name} & \colhead{$r_{39}(\degree)$} & \colhead{$TS$} & \multicolumn{1}{c}{$\Gamma_{\rm WCDA}$} & \multicolumn{1}{c|}{$\Gamma_{\rm KM2A}$} & \colhead{Name} & \colhead{$\Gamma$} & \colhead{$TS$} & \colhead{ASSOC\_GAM} & \colhead{CLASS} & \colhead{ASSOC1}  
}   
\startdata
\hline
  J0343+5254* & $0.33\pm0.05$ & 94.1    & $1.7\pm0.19$  & $3.5\pm0.1$   & 2FLE J0340.8+5258 & $1.95\pm0.22$ & 161.87 & 4FGL J0340.4+5302 &  &    \\
\hline
  J0703+1405 & $1.3 \pm0.21$  & 90.2     & $1.98\pm0.12$ & $3.98\pm0.08$  & 2FLE J0658.7+1420 & $1.79\pm0.29$ & 217.94 & 4FGL J0659.7+1416   & PSR & PSR J0659+1414 \\
  J1954+3253 & $0.17\pm0.03$ & 144.0  & $2.45\pm0.09$  & --  & 2FLE J1954.5+3314 & $2.11\pm0.43$ & 72.54   & 4FGL J1954.4+3252e & SPP & CTB 80 \\
  J1956+2921 & $0.78\pm0.05$ & 151.3 & $2.03\pm0.06$  & $3.42\pm0.12$ & 2FLE J1955.7+2838 & $0.73\pm0.96$ & 45.45   & 4FGL J1954.3+2836   & PSR & PSR J1954+2836 \\
  J2028+3352 & $1.7\pm0.23$   & 75.7    & -- & $3.38\pm0.19$ & 2FLE J2025.5+3335 & $0.64\pm1.90$ & 26.53   & 4FGL J2025.3+3341   & bcu & PSR J2028+3332\tablenotemark{c} \\
  J2238+5900 & $0.51\pm0.04$ & 110.2 & $2.39\pm0.07$ & $3.55\pm0.09$ & 2FLE J2236.6+5841 & $0.98\pm1.25$ & 25.18   & 4FGL J2238.5+5903   & PSR & PSR J2238+5903 \\
\hline
\enddata
\footnotesize
\tablenotetext{a}{Source properties reported in the LHAASO catalog \citep{LHAASO}: source name (* indicates that the detected source suffers from Galactic gamma-ray emission significantly in testing);  39\% uncertainty radius for size of extended source ($r_{39}$); photon index ($\Gamma$); significance of detection ($TS$). }
\tablenotetext{b}{2FLE source properties reported in the 2FLE catalog (see Table~\ref{tab:cat_col_description} for column description). }
\tablenotetext{c}{Note that the 4FGL associated counterpart (ASSOC\_2) is B2 2023+33 (see Section~\ref{sec:source_assoc} for details). }
\end{deluxetable*}\label{tab:lhaaso_2fle_comparison}

\subsection{Comparison with 1FLE}\label{sec:1fle_comp}
Using 8 years and 9 months of Fermi-LAT data and the PGWave wavelet method for source detection, the 1FLE catalog reported 198 sources detected at $>3\sigma$ in the $31.6-100$ MeV energy range, 14 of which had been previously identified by COMPTEL (\citealp{1fle_principe}). Similar to the 1FLE, we simplify these ranges by stating $30-100$ MeV to mean $31.6-100$ MeV. 
Compared to the analysis performed in the 1FLE, the exposure time used in the 2FLE is 1.6 times larger, and the energy range considered is wider.  
Taken together, these factors provide an increase in photon statistics of about 2.18, and hence a $\sim1.48$ higher sensitivity. Moreover, we used a maximum likelihood method for source finding, we included the Galactic diffuse template and we enabled energy dispersion. Further, the higher maximum energy adopted by the 2FLE allows to benefit from a narrower PSF at energies above 100 MeV. Collectively, these differences provide the ability to detect more sources ($\times 2.2$) at a higher significance detection threshold ($>5\sigma$) and at a deeper flux limit ($\tilde{F}_{\rm 30-100\, MeV, 2FLE}\sim \tilde{F}_{\rm 30-100\, MeV, 1FLE}/1.6$).

We perform a positional crossmatch between the 1FLE and the 2FLE sources with TopCAT \texttt{SkyEllipse}, using as positional uncertainties a flat $1.5\degree$ radius for the 1FLE\footnote{The 1FLE utilized a flat $1.5\degree$ radius when searching for associations to the \textit{Fermi} Large Area Telescope Third Source Catalog (3FGL, \citealp{3FGL}).} and the \texttt{Conf\_95\_SemiMajor}, \texttt{Conf\_95\_SemiMinor} and \texttt{Conf\_95\_PosAng} listed in the 2FLE as semimajor, semiminor and angle of the ellipse. This match returns 166 1FLE sources ($84\%$ of the 1FLE catalog) that are spatially coincident with a 2FLE one. Of the 166 positional coincidences: 17 2FLE sources do not have a 4FGL counterpart; 3 of the matching 1FLE sources do not have a 3FGL counterpart; of the 147 that have both a 4FGL (2FLE) and a 3FGL (1FLE) counterpart, 124 have the same $\gamma$-ray counterpart, while 23 2FLE $\gamma$-ray counterparts (4FGL names) are different from the 1FLE $\gamma$-ray counterparts (3FGL names).

\begin{figure}
    \centering
    \includegraphics[width=\columnwidth]{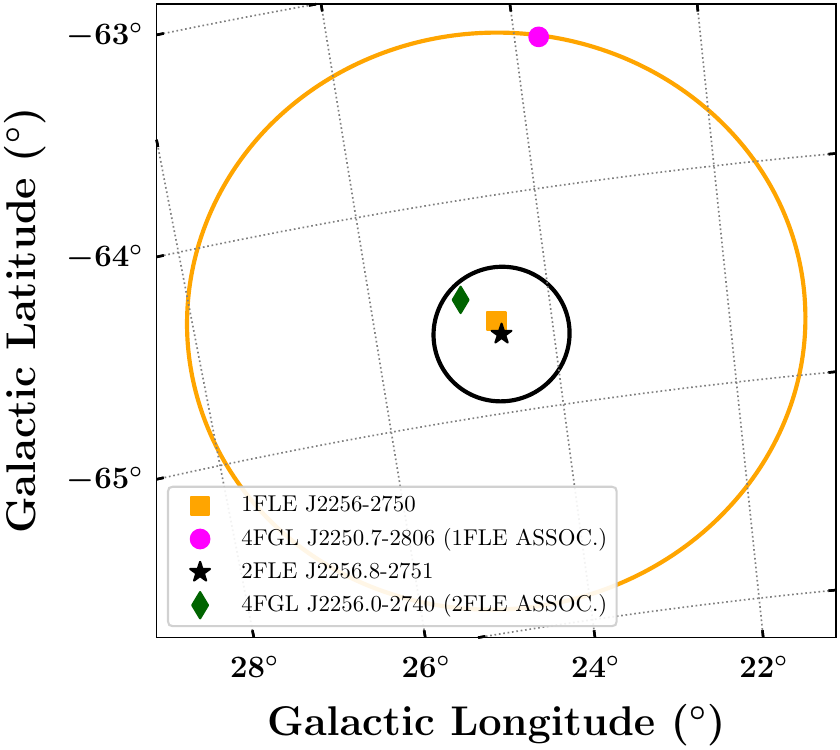}
    \caption{Field of view with position and 95\% positional uncertainty of 2FLE J2256.8-2751 (black star), 2FLE associated 4FGL source position and 95\% positional uncertainty (green diamond), the 1FLE location position and $1.5\degree$ positional uncertainty (orange square), and 1FLE associated 4FGL  source position and 95\% positional uncertainty (magenta circle). The positional uncertainties of the 4FGL counterparts are very small, and therefore are not visible in this plot. In this instance, the 2FLE association is most likely the true one based on location alone.}
    \label{fig:2fle_1fle_assoc_battle}
\end{figure}

The majority (16/17) of  sources that lack a $\gamma$-ray counterpart in the 2FLE are associated to a nearby $\gamma$-ray source in the 1FLE. This difference is mostly explained by the larger positional uncertainty applied to the 1FLE while searching for counterparts than the 2FLE, which reports stringent constraints on the source localization, and to blending effects affecting the 2FLE localization (see Sections~\ref{sec:source_assoc} and~\ref{sec:unassoc}). It is likely that these 1FLE sources are indeed the same source detected in the 2FLE. The disagreement in $\gamma$-ray counterparts for 23 sources is attributed both to the same localization difference as well as the depth of the 4FGL with respect to the 3FGL. For example, the source 1FLE J0536-6856 is associated to the LMC and is close to PSR J0540-6919 and N 157B (toward the LMC center). The closest 2FLE source is 2FLE J0542.4-6919, associated explicitly to PSR J0540-6919 (but on the other side) which suggests that the  
1FLE localization suffered from source confusion between the LMC and the pulsar.
On the other hand, the source 2FLE J2256.8-2751 is associated with the FSRQ PKS 2253-278 (4FGL J2256.0-2740) which was not detected by the 3FGL; the corresponding 1FLE J2256-2750 is associated with the BCU PMN J2250-2806 (3FGL/4FGL J2250.7-2806) located $1.5^{\circ}$ away from the 1FLE position (see Figure~\ref{fig:2fle_1fle_assoc_battle}, left panel).

Of the remainder 32 1FLE sources, we find that 17 are positionally coincident with 2FLE sources with $TS<25$ and thus are not reported in the final 2FLE catalog (see Section~\ref{sec:pipeline}). These sources may have lower significance in our analysis because of source variability or the different way the background is taken into account in the 1FLE and 2FLE analyses. Moreover, 3/32 match a 2FLE source removed from the 2FLE main catalog 
because it was found close to an ROI boundary and not confirmed in the neighboring ROI covering that location better (Section~\ref{sec:pipeline}). Furthermore, six out of the 12 remaining 1FLE sources not found in 2FLE are at low Galactic latitudes $|b|<5\deg$, where the Galactic diffuse emission plays a major role in the source’s detectability. Finally, for the last 6 1FLE sources ($|b|>5\deg$; 1FLE J2206+7040, 1FLE J1138+3832, 1FLE J0424+1446, 1FLE J0548-0050, 1FLE J1033+1601, 1FLE J1720+0916), we find that four (1FLE J1138+3832, 1FLE J0424+1446, 1FLE J0548-0050, 1FLE J1033+1601) are not spatially coincident with any 4FGL-DR4 source within their $1.5\degree$ association radius. Considering the location of these sources and blending effects affecting both catalogs, the fact that no MeV excess was found at those locations by the 2FLE is not cause for concern.

\subsection{Comparison with AGILE}
Another instrument that has some sensitivity in the MeV band is the AGILE Gamma-Ray Imager Detector (GRID). The Second AGILE catalog of gamma-ray sources \citep[2AGL,][]{2agl} reports 31 sources detected above $>4\sigma$ in the 30--100 MeV band, the majority of which (70\%) lie within the Galactic plane ($|b|<5\degree$). For completeness, we cross-matched the 2FLE with the 2AGL
with TopCAT \texttt{SkyEllipse} using the $\gamma$-ray coordinates and positional uncertainties for the 2AGL and we found 16 matches. 
The rest of the 2AGL sources for which we do not find a 2FLE counterpart all lie within the Galactic plane, the region where these catalogs suffer the most from source confusion and blending (as described in Section~\ref{sec:1fle_comp}), hence this mismatch is not unexpected.

\subsection{20-200 MeV Band Spectral Constraints}\label{sec:sed_constraints}

The 2FLE contains 387 sources associated with a known 4FGL source. Of these, in the 4FGL-DR4 47 are modeled as a \texttt{PowerLaw},  300  with \texttt{LogParabola} and 40 with \texttt{PLSuperExpCutoff4}. We look for sources that may have a spectral break between the 2FLE and 4FGL band. We do so by comparing the primary spectral index parameter for each model to \texttt{PowerLaw\_Index} from the 2FLE. For \texttt{PowerLaw} models we used \texttt{PL\_Index}. In the case of sources best fit with \texttt{LogParabola} and \texttt{PLSuperExpCutoff4}, we used \texttt{LP\_Index} and \texttt{PLEC\_IndexS}, respectively, propagated to their value at $E=100$\, MeV.  We identify potential spectral breaks when the 2FLE and 4FGL indices are different at 2$\sigma$ (see Figure~\ref{fig:2fle_index_vs_4fgl_src_compare}) using the following criterion:
\begin{equation}
    \frac{|\Gamma_{\rm 2FLE}-\Gamma_{\rm 4FGL}|}{\sqrt{\Gamma_{\rm err, 2FLE}^2+\Gamma_{\rm err, 4FGL}^2}} > 2.
\end{equation}

We found a total of 92 potential spectral breaks. 
Considering the boundary for a transition of a spectrum from soft to hard to be $\Gamma=2$, we find that 59/92 display a hard 2FLE spectral index and a soft 4FGL one, indicative of the fact that the 2FLE may be sampling close to a peak of the emission (in a $\nu F_{\nu}$ plot) and capturing the rising edge of the spectrum as the energy increases. Interestingly, 70\% of the sources of Galactic origin show this hard-to-soft spectral behavior (Figure~\ref{fig:2fle_index_vs_4fgl_src_compare}, right panel), compared to only 30\% of the extragalactic sample. On the other hand, we also have 9/92 sources displaying the opposite behavior (soft 2FLE spectrum and hard 4FGL one). However, this number is compatible with statistical fluctuations since at the $2\sigma$ level 2.3\% of cases are expected to be harder in the 2FLE than the 4FGL-DR4 just by chance. This break could indicate that the 2FLE and 4FGL spectra are sampling two different components of the sources' physical SED (e.g.~a transition from a synchrotron to an inverse Compton emission). Lastly, we find 21/92 (3/92) cases in which the spectrum is soft (hard) in both catalogs, but discrepant $>2\sigma$; these cases could indicate an intrinsic spectral curvature in the sources' $\gamma$-ray emission.

We also checked if variability could be playing a role in these spectral changes. Of the 387 4FGL-DR4 sources associated with the 2FLE objetcs, 83\% have a \texttt{Variability\_Index}  $>27.69$ (i.e.~are significantly variable in the 4FGL band). About $\sim66\%$ of those sources that exhibit a break are variable, hence we cannot exclude that some of these differences in spectral indices could be attributed to variability. However, to validate this assumption, we would need to compute the light curves in the MeV band, and correlate it to the higher energy bands, something we leave for future work.

\begin{figure*}[h]
    \centering
    \includegraphics[width=0.48\textwidth]{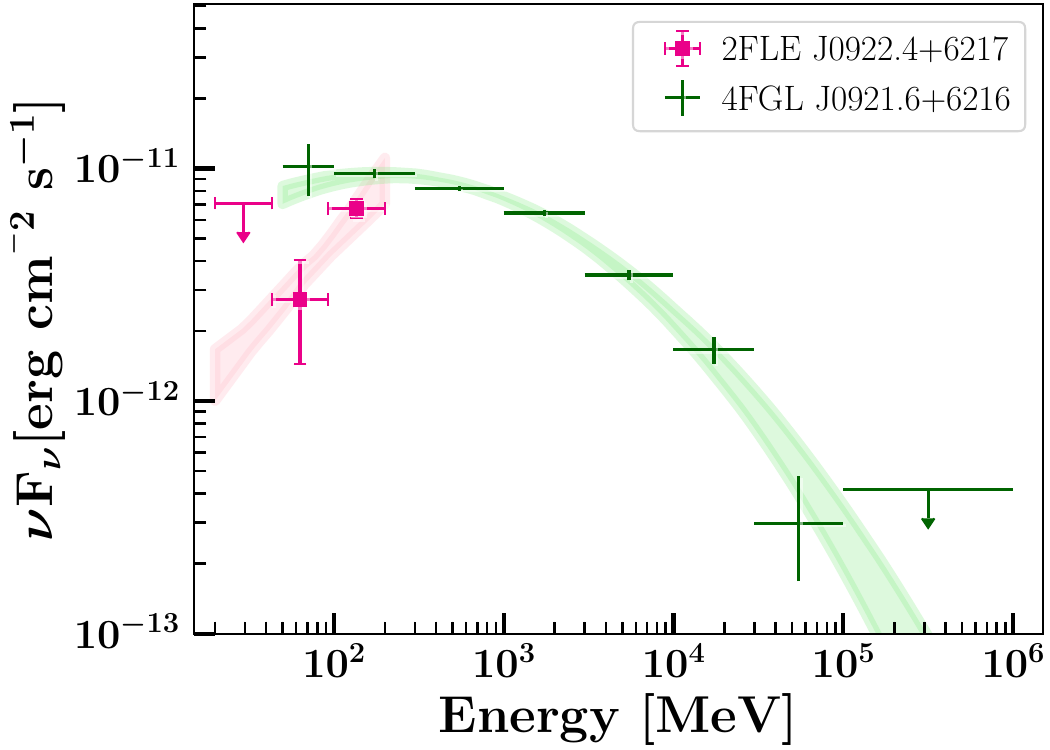}
    \includegraphics[width=0.48\textwidth]{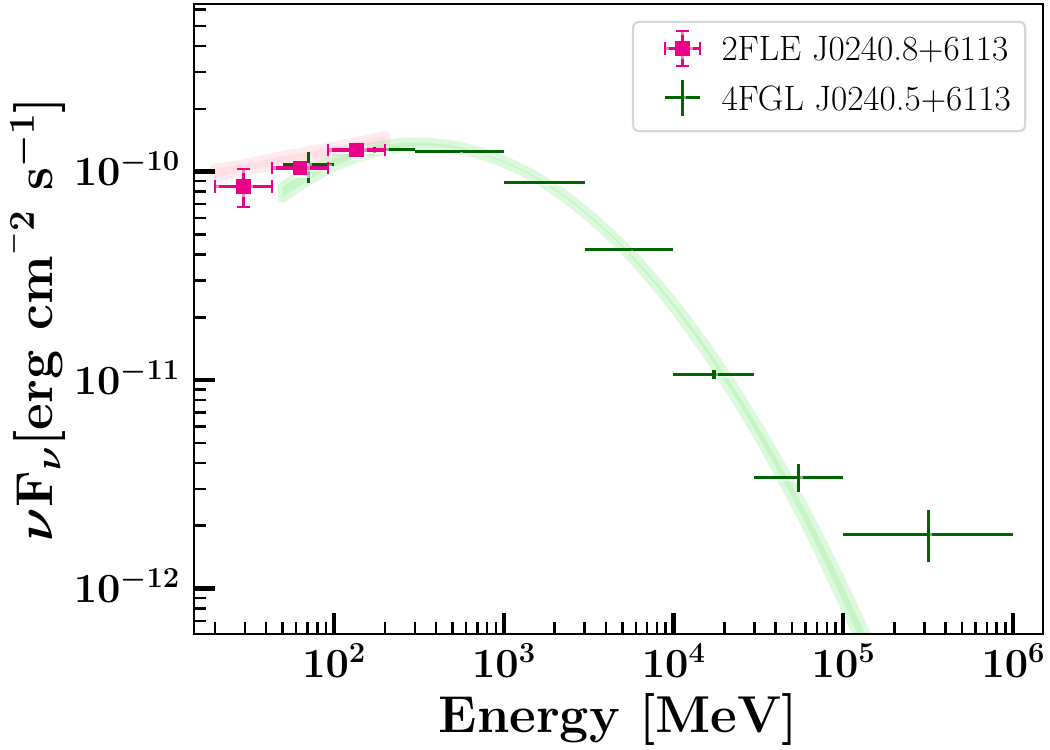}
    \includegraphics[width=0.48\textwidth]{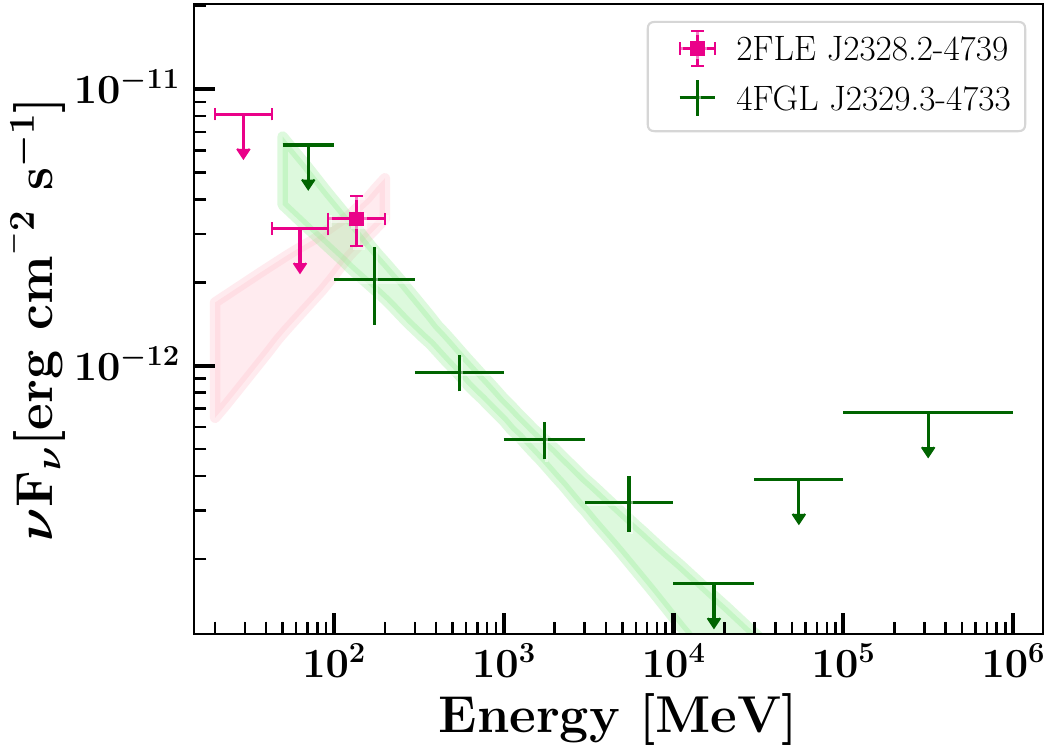}
    \includegraphics[width=0.48\textwidth]{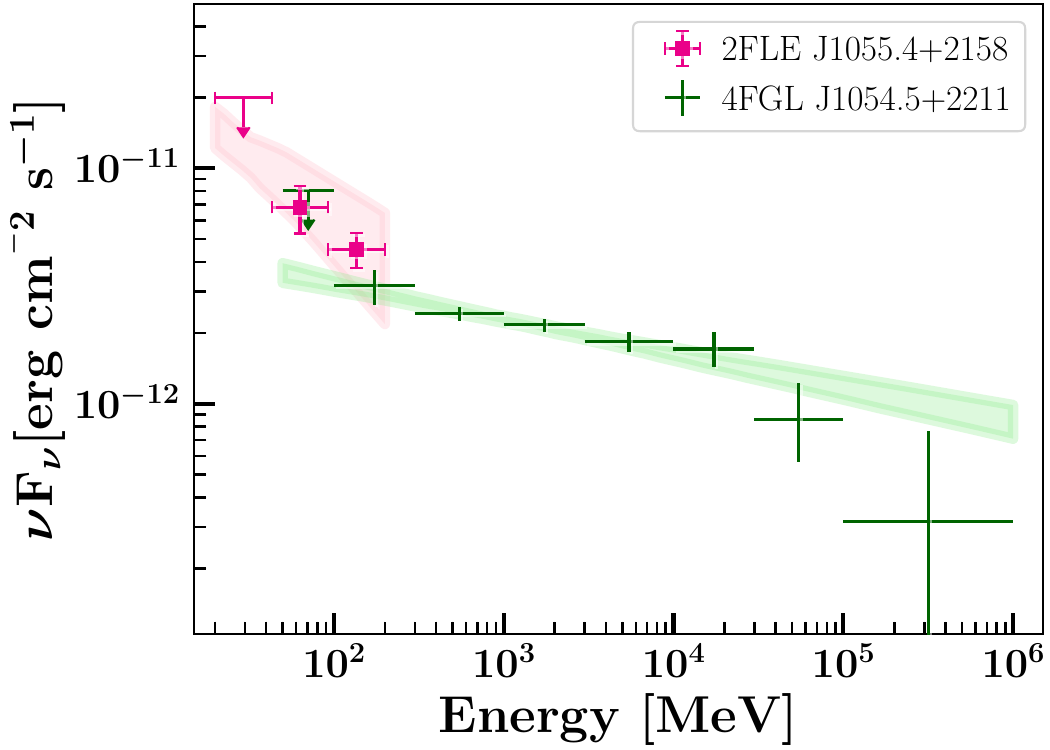}
    \caption{Combined 2FLE (pink) and 4FGL (green) spectral energy distributions. The shaded bands show the $1\sigma$ confidence levels obtained using the best-fit parameters for the sources. Top Left: 2FLE J0922.4+6217 associated with FSRQ OK 630 depicts how the 2FLE energy range constrains the location of the peak emission. Top right: HMB LSI +61 303, with the 2FLE we see the peak is flatter than predicted by the 4FGL model. Bottom left: FSRQ PKS 2326-477 modeled by a power law in the 4FGL. Similar to the example above, the 2FLE enables us to determine the peak of the emission and see the turnover of the slope of the gamma-ray emission. Bottom right: BLL  associated with 87GB 105148.6+222705. Here we see that an extrapolation of the 4FGL model would predict a much less steep slope than what is found by the 2FLE. }  
    \label{fig:2fle_4fgl_sed_benefit}
\end{figure*}

Figure \ref{fig:2fle_4fgl_sed_benefit} displays 4 sources with their 2FLE SED data points and best fit model in the 20-200 MeV energy range (magenta) and the best fit model reported by the 4FGL along with its SED data points from the 50 MeV - 1 TeV energy range (green). 
We selected these cases to demonstrate the benefit of the 2FLE for both Galactic and extragalactic sources, as well as sources modeled with and without a power law. Furthermore, we see in the plots for 2FLE J0922.4+6217 and 2FLE J2328.2-4739 how the 2FLE can constrain the peak emission of the source under investigation when the MeV gamma ray behavior has a spectral break from that of the GeV. Lastly, 2FLE J1055.4+2158 is an interesting case where the 20-200 MeV data show that the SED may be much softer at MeV energies. As part of this catalog we have generated similar SEDs for all the 2FLE sources (Section~\ref{sec:sed_gen}) with well constrained spectral parameters (i.e.~Flux\_Flag=0) associated with a 4FGL source and have provided them in the official 2FLE website\footnote{\url{https://fermi.gsfc.nasa.gov/ssc/data/access/lat/2FLE/}}.

\section{Summary and Conclusions}\label{sec:summary}
We performed an all-sky analysis of the 20-200 MeV low-energy gamma-ray sky based on 14 years of the \textit{Fermi}-LAT mission. This catalog seeks to reduce the ``MeV gap", the poorly charted energy regime stretching between X-rays and GeV energies, 
by leveraging the most sensitive instrument currently available. We present the 2FLE catalog containing 450 sources detected at $TS>25$ ($\sim5\sigma$). The 2FLE contains just over twice as many sources than its predecessor, the 1FLE. 
In general, the 2FLE contains fainter sources, more significantly detected, and better constrained in flux than reported in the 1FLE.

Of the 450 sources detected, 377 are associated to previously known sources, with 86\% of these (325/377) of extragalactic origin and 14\% (52/377) of Galactic origin. The dominant source class represented are flat-spectrum radio quasars (FSRQs) representing 58\% (218/377) of the associated sources. Galactic associated sources are distinguished from extragalactic sources in the 2FLE by their characteristically harder spectrum ($\Gamma_{\rm GAL} \sim 1.48$, $\Gamma_{\rm EGAL}\sim1.99$). The Galactic source classes are primarily represented by pulsars, millisecond pulsars, and supernova remnant/pulsar wind nebulae. Of the associated sources in the 2FLE, 3
are not detected in the 4FGL-DR4, one FSRQ, one BCU and one globular cluster (M2) detected for the first time in gamma rays.

There remain 73 unassociated sources reported in the 2FLE with 41 sources detected with a $TS>100$. The distribution of the unassociated sources' photon index and Galactic latitude more closely aligns with sources of extragalactic origin ($\Gamma_{\rm UNASSOC}\sim 2$). Of the 73 unassociated sources, 16 are found at low latitudes ($|b|<10\degree$). Considering the prevalence of Galactic sources at these latitudes, future studies and  multiwavelength campaigns can help to determine the origin of these low-latitude unassociated detections.

Furthermore, we find that 6 2FLE sources lacking a TeV counterpart are found spatially coincident with sources reported in the First LHAASO Catalog of Gamma-ray Sources, one of which lacking any associations.

Lastly, a major benefit of the 2FLE has been probing lower into the MeV regime, where many sources are expected to peak in their emission. By providing the best-fit spectral models in the 20-200 MeV energy range we further constrain the low-energy gamma ray behavior of these detected sources. Sources detected in the 2FLE are at (or near) their SED peak, rendering the 20-200 MeV band an important one to characterize the energetics of gamma-ray sources.

The 2FLE is the largest and most sensitive catalog of the MeV sky to date. Moreover, thanks to more than 20 years of research and development, another portion of the MeV gap will soon be probed between  0.2 -- 5 MeV by the Compton Spectrometer and Imager (COSI; \citealp{COSI}), scheduled to launch in 2027. Additionally, there are a number of larger mission concepts currently being proposed such as AMEGO-X (\citealp{AMEGOX}), e/new-ASTROGAM (\citealt{tatischeff2016astrogam,e2017astrogam,newASTROGAM}) and GECCO (\citealp{GECCO}). These future missions seek to transcend both COSI's and \textit{Fermi}-LAT's capabilities and contribute to comprehensively observing the long-standing MeV Gap. 

\software{
    Astropy \citep[\hspace{-0.2cm}][]{astropy_i,astropy_ii,astropy_iii},
    numpy (\citealp{Numpy}),
    SAO Image DS9 (\citealp{DS9}),
    TOPCAT (\citealp{topcat}),
    fermitools, \textit{fermipy}
}

\begin{contribution}
{\bf S. Joffre \& L. Marcotulli}: Leading the core scientific analysis and result interpretation. Writing - original draft. 
{\bf M. Ajello}: Supervising and project planning, assisting in the data analysis and the interpretation of results. Writing - review \& editing.
{\bf C. Karwin}: Leading the diffuse emission models and systematic analysis, and contributing to the interpretation of results. Writing - review \& editing.
{\bf B. Lott \& D. Gasparrini}: Responsible for the multiwavelength counterpart association. Writing - review \& editing.
\end{contribution}

\begin{acknowledgments}
The Fermi LAT Collaboration acknowledges generous ongoing support from a number of agencies and institutes that have supported both the development and the operation of the LAT as well as scientific data analysis. These include the National Aeronautics and Space Administration (NASA) and the Department of Energy in the United States, the Commissariat {\`a} l’Energie Atomique and the Centre National de la Recherche Scientifique / Institut National de Physique Nucl\'{e}aire et de
Physique des Particules in France, the Agenzia Spaziale Italiana and the Istituto Nazionale di Fisica Nucleare in Italy, the Ministry of Education, Culture, Sports, Science and Technology (MEXT), High Energy Accelerator Research Organization (KEK) and Japan Aerospace Exploration Agency (JAXA) in Japan, and the K. A. Wallenberg Foundation, the Swedish Research Council and the Swedish National Space Board in Sweden, and the IceCube Collaboration. 

Specifically we would like to thank NASA for funding from the \textit{Fermi}-LAT Guest Investigator Proposal Number 141104 and 161072  that helped to fund this work. This work was performed in part under DOE Contract DE- AC02-76SF00515.
Additional support for science analysis during the operations phase is gratefully acknowledged from the Istituto Nazionale di Astrofisica in Italy and the Centre National d’Etudes Spatiales in France.

C.M.K.’s research was supported by an appointment to the NASA Postdoctoral Program at NASA Goddard Space Flight Center, administered by Oak Ridge Associated Universities under contract with NASA.

LM acknowledges that this work was supported by the Initiative and Networking Fund of the Helmholtz Association under the Helmholtz Investigator Groups Programme, call 2025 (VH-NG-21-01). LM acknowledges support from DESY (Zeuthen, Germany), a member of the Helmholtz Association HGF.

M.A. acknowledges the NASA contracts 80NSSC23K1491 and 80NSSC21K1914 for this project. 

\end{acknowledgments}

\appendix
\section{Description of the FITS Version of the 2FLE Catalog}\label{app:a}
The FITS format version of the 2FLE catalog has two binary table extensions. The extension \texttt{2FLE\_point\_source\_catalog} contains all of information about the sources and its format is described in Table~\ref{tab:cat_col_description}. The extension \texttt{ID\_CAT\_REFERENCE} has all the information of the catalogs used for the association procedure described in Section~\ref{sec:source_assoc}, and its columns are described in Table~\ref{tab:fits_extension}.

\renewcommand{\arraystretch}{1} 
\setlength{\tabcolsep}{0.95pt}  
\begin{deluxetable}{cc}[h]
\centering
\tablecaption{Description of the 2FLE FITS extension \texttt{ID\_CAT\_REFERENCE}}
\tablehead{ \colhead{Column}  & \colhead{Description} } 
\startdata
ID\_Catalog & Numerical ID for the catalog \\
Name & Official name of the catalog \\
Reference & Reference paper of the catalog  \\
URL & Link to the catalog \\
Prefix & Prefix for the catalog\\
\enddata
\end{deluxetable}\label{tab:fits_extension}

\clearpage
\newpage
\bibliography{mev_biblio}{}
\bibliographystyle{aasjournalv7}

\end{document}